\documentclass[12pt]{article}
\usepackage{epsfig,amsmath}
\begin{document}
%\@addtoreset{equation}{section}
\renewcommand{\theequation}{\arabic{section}.\arabic{equation}}%

\newcounter{saveeqn}%
\newcommand{\alpheqn}{\setcounter{saveeqn}{\value{equation}}%
\stepcounter{saveeqn}\setcounter{equation}{0}%
\renewcommand{\theequation}
       {\mbox{\arabic{section}.\arabic{saveeqn}\alph{equation}}}}%
\newcommand{\reseteqn}{\setcounter{equation}{\value{saveeqn}}%
\renewcommand{\theequation}{\arabic{section}.\arabic{equation}}}%

%\begin{document}
\title{\hfill OKHEP--99--01\\
The Casimir Effect:\\Physical Manifestations of Zero-Point 
Energy\thanks{Invited Lectures at the 17th Symposium on
Theoretical Physics, Seoul National University, Korea, June 29--July 1, 1998.}}
\author{KIMBALL A. MILTON\\
Department of Physics and Astronomy,\\ The University of Oklahoma\\
Norman, OK 73019-0225 USA}
\maketitle

\def\bGamma{\mbox{\boldmath{$\Gamma$}}}
\def\bPhi{\mbox{\boldmath{$\Phi$}}}
\def\bnabla{\mbox{\boldmath{$\nabla$}}}
\begin{abstract}Zero-point fluctuations in quantum fields give rise to
observable forces between material bodies, the so-called Casimir forces.  
In these lectures I present the theory of the Casimir effect, primarily
formulated in terms of Green's functions. There is an intimate relation
between the Casimir effect and van der Waals forces.
Applications to conductors
and dielectric bodies of various shapes will be given for the cases of
scalar, electromagnetic, and fermionic fields.  The dimensional dependence 
of the effect will be described. Finally, we ask the question: Is there
a connection between the Casimir effect and the phenomenon of
sonoluminescence?
\end{abstract}

\section{Introduction}
The Casimir effect is a manifestation of macroscopic quantum field theory.
Its origin is intimately tied to the van der Waals forces between neutral
molecules.  In 1873 van der Waals \cite{Waals} introduced these weak 
intermolecular forces to explain derivations from the ideal gas laws, but
their physical basis did not begin to emerge until 1930 when London 
\cite{London} showed how the leading behavior of this force could be 
understood from quantum mechanics.  That is, the Hamiltonian describing
the interaction between two dipoles separated by a distance $R$ is
\begin{equation}
H_{\rm int}={{\bf d}_1\cdot{\bf d}_2-3{\bf d}_1\cdot{\bf n}\,{\bf d}_2\cdot
{\bf n}\over R^3},
\end{equation}
where $\bf n$ is the relative direction of the two dipoles.  Because
$\langle {\bf d}_1\rangle=\langle {\bf d}_2\rangle=0$, this Hamiltonian
has vanishing expectation value, $\langle H_{\rm int}\rangle=0$;
however, in second order of perturbation theory this operator appears squared,
so there is an effective interaction potential between the dipoles,
\begin{equation}
V_{\rm eff}\sim{1\over R^6}.
\label{london}
\end{equation}
This is a short distance electrostatic effect.
The next step in the saga occurred in 1948 when Casimir and Polder 
\cite{CasimirandPolder} introduced {\em retardation}, resulting in a potential
which for large distances falls off faster,
\begin{equation}
V_{\rm eff}\sim {1\over R^7}.
\label{retdis}
\end{equation}
This result can be understood by dimensional considerations.  For sufficiently
weak fields, the polarization of a molecule should be proportional to the
electric field acting on the molecule,
\begin{equation}
{\bf d=\alpha E},
\end{equation}
where $\alpha$ is the polarizability of the molecule.  Each atom produces
a dipole field, so the two atoms polarize each other.  In terms of a unit
of length $L$, we have dimensionally
\begin{equation}
[d]=eL,\quad [E]=eL^{-2}, \quad \mbox{so}\quad [\alpha]=L^3.
\end{equation}
Thus we have, at zero temperature,
\begin{equation}
V_{\rm eff}\sim {\alpha_1\alpha_2\over R^6}{\hbar c\over R}.
\end{equation}
[It may be useful to remember the conversion factor $\hbar c\approx
2\times 10^{-5}$ eV cm.]
The high temperature limit is classical,
\begin{equation}
V_{\rm eff}\sim{\alpha_1\alpha_2\over R^6}kT, \quad T\to\infty.
\end{equation}

Here we have thought of the interactions from an action-at-a-distance
point of view.  From this viewpoint, the Casimir effect is simply
a macroscopic manifestation of the sum of a multitude of molecular
van der Waals interactions.  However, shortly after the original
Casimir and Polder paper, Bohr suggested to Casimir that zero-point
energy was responsible for the intermolecular
force, and indeed that is an equivalent
viewpoint \cite{CasimirLeipzig}. One can shift the emphasis from action
at a distance between molecules to local action of fields.

The connection between the sum of the zero-point energies of the modes
and the vacuum expectation value of the field energy may be easily given.
Let us regulate the former with an oscillating exponential:
\begin{eqnarray}
{1\over2}\sum_a\hbar\omega_a e^{-i\omega_a\tau}
={\hbar\over2}\int_{-\infty}^\infty {d\omega\over2\pi i}e^{-i\omega\tau}\omega
\sum_a{2\omega\over\omega_a^2-\omega^2-i\epsilon},\quad\tau\to0,
\label{zeropt}
\end{eqnarray}
where $a$ labels the modes, and the contour of integration in the second form
may be closed in the lower half plane.   For simplicity of notation let us
suppose we are dealing with massless 
scalar modes, for which the eigenfunctions and eigenvalues satisfy
\begin{equation}
-\nabla^2\psi_a=\omega_a^2\psi_a.
\end{equation}
Because these are presumed normalized, we may write the second form in 
Eq.~(\ref{zeropt}) as 
\begin{eqnarray}
&&\hbar\int{d\omega\over2\pi i}\omega^2 e^{-i\omega\tau}\int(d{\bf x})
\sum_a{\psi_a({\bf x})\psi_a^*({\bf x})\over\omega_a^2-\omega^2-i\epsilon}
\nonumber\\
&&\quad
={\hbar\over i}\int(d{\bf x})
\int{d\omega\over2\pi}\omega^2G({\bf x,x};\omega)e^{-i\omega
(t-t')}\bigg|_{t\to t'}\nonumber\\
&&\quad=\int(d{\bf x})
\partial^0\partial^{\prime0}\langle\phi(x)\phi(x')\rangle\big|_{x'\to x},
\end{eqnarray}
where the Green's function $G({\bf x},t;{\bf x'},t')$ satisfies 
\begin{equation}
\left(-\nabla^2+{\partial^2\over\partial t^2}\right)G({\bf x},t;{\bf x'},t')
=\delta({\bf x-x'})\delta(t-t'),
\end{equation}
and is related to the vacuum expectation value of the time-ordered
product of fields according to
\begin{equation}
G({\bf x},t;{\bf x'},t')={i\over\hbar}\langle{\rm T}
\phi({\bf x},t)\phi({\bf x'},t')\rangle.
\label{vevgreen}
\end{equation}
For a massless scalar field, the canonical energy-momentum tensor is
\begin{equation}
T^{\mu\nu}=\partial^\mu\phi\partial^\nu\phi-{1\over2}g^{\mu\nu}\partial^\lambda
\phi\partial_\lambda\phi.
\label{stresstensor}
\end{equation}
The second term involving the Lagrangian in Eq.~(\ref{stresstensor}) may be 
easily shown not to contribute when integrated over all space, by virtue of
the equation of motion, $-\partial^2\phi=0$ outside the sources,
 so we have the result, identifying the
zero-point energy with the vacuum expectation value of the field energy,
\begin{equation}
{1\over2}\sum_a\hbar\omega_a=\int(d{\bf x})\langle T^{00}({\bf x})\rangle.
\end{equation}
In the vacuum this is divergent and meaningless.  What is observable is the
{\em change\/} in the zero-point energy when matter is introduced.  In this
way we can calculate the Casimir forces.

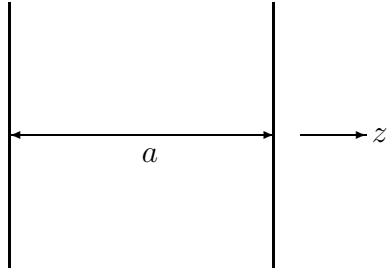
\begin{figure}[ht]
\centering
\begin{picture}(200,100)
\thicklines
\put(50,0){\line(0,1){100}}
\put(150,0){\line(0,1){100}}
\thinlines
\put(100,50){\vector(1,0){50}}
\put(100,50){\vector(-1,0){50}}
\put(100,40){$a$}
\put(160,50){\vector(1,0){25}}
\put(190,50){\makebox(0,0){$z$}}
\end{picture}
\caption{Geometry of parallel conducting plates.}
\label{fig1}
\end{figure}

Alternatively, we can calculate the stress on the material bodies.
Figure \ref{fig1} shows the original geometry considered by Casimir,
where he calculated the quantum fluctuation force between parallel,
perfectly conducting plates \cite{Cas}.
The force per unit area $f$ on one of the plates is given in terms of the
normal-normal component of the stress tensor,
\begin{equation}
f=\int dx\,dy\,\langle T_{zz}\rangle,
\end{equation}
where the integral is over the surface area of the plate in question.
For electromagnetic fields, the relevant stress tensor component is
\begin{equation}
T_{zz}={1\over2}(H_\perp^2-H_z^2+E_\perp^2-E_z^2).
\end{equation}
We impose classical boundary conditions on the surfaces,
\begin{equation}
H_z=0,\quad {\bf E}_\perp=0,
\end{equation}
and the calculation of the vacuum expectation value of the field components
reduces to finding the classical TE and TM Green's functions.  In general,
one further has to subtract off the stress that the formalism would give
if the plates were not present, the so-called volume stress,
and then the result of a simple calculation, which is given below,
is
\begin{equation}
f=\int dx\,dy\left[T_{zz}-T_{zz}({\rm vol})\right]=-{\pi^2\over 240 a^4}
\hbar c,
\end{equation}
an attractive force.

The dependence on $a$ is, of course, completely determined by dimensional
considerations.  Numerically, the result is quite small,
\begin{equation}
f=-8.11\,{\rm MeV}\,{\rm fm}\,a^{-4}=-1.30\times 10^{-27} {\rm N}\,{\rm m}^2
\, a^{-4},
\end{equation}
and will be overwhelmed by electrostatic repulsion between the plates
if each plate has an excess electron density $n$ greater than $1/a^2$,
from which it is clear that the experiment must be performed at the
$\mu$m level.
Nevertheless, it has been directly measured to
an accuracy of several percent 
\cite{deriagin,derjaguin,kitchener,sparnaay,black,silfhout,tabor0,tabor,winterton,israelachivili}.  
(The cited measurements include insulators as well as conducting surfaces;
the corresponding theory will be given in Section 3.)
Until recently, the most convincing experimental evidence comes
from the study of thin helium films \cite{sabisky}; there the
corresponding Lifshitz theory \cite{lifshitz,dzyaloshinskii}
has been confirmed over nearly
 5 orders of magnitude in the van der Waals potential
(nearly two orders of magnitude in distance).  Quite recently, the Casimir 
effect between conductors has been confirmed to the 5\% level by
Lamoreaux \cite{lamoreaux}, and to 1\% by Mohideen and Roy \cite{roy}.

\section{Casimir Force Between Parallel Plates}
\setcounter{equation}{0}%
\label{sec:parallel}
\subsection{Dimensional Regularization}
\label{dimreg}
We begin by presenting a simple, ``modern,'' derivation of the Casimir
effect in its original context, the electromagnetic force between
parallel, uncharged, perfectly conducting plates.  No attempt at
rigor will be given, for the same formul\ae\ will be derived by
a consistent Green's function technique in the following subsection.
Nevertheless, the procedure illustrated here correctly produces
the finite, observable force starting from a divergent formal expression,
without any explicit subtractions, and is therefore of great utility
in practice.

For simplicity we consider a massless scalar field $\phi$
confined between two parallel
plates separated by a distance $a$.  (See Fig.\ \ref{fig1}.)
Assume the field satisfies Dirichlet boundary conditions on the plates,
that is 
\begin{equation}
\phi(0)=\phi(a)=0.
\label{eq:bc}
\end{equation}
The Casimir force between the plates results from the zero-point energy
per unit transverse area
\begin{equation}
u={1\over2}\sum \hbar\omega={1\over2}\sum_{n=1}^\infty\int{d^2k\over(2\pi)^2}
\sqrt{k^2+{n^2\pi^2\over a^2}},
\label{zeropt2}
\end{equation}
where we have set $\hbar=c=1$, and introduced normal modes labeled by
the positive integer $n$ and the transverse momentum $k$.
 
To evaluate Eq.~(\ref{zeropt2}) we employ dimensional regularization.
That is, we let the transverse dimension be $d$, which we will subsequently
treat as a continuous, complex variable.  It is also convenient to
employ the Schwinger proper-time representation for the square root:
\begin{equation}
u={1\over2}\sum_n\int{d^dk\over(2\pi)^d}\int_0^\infty{dt\over t}t^{-1/2}
e^{-t(k^2+n^2\pi^2/a^2)}{1\over\Gamma(-{1\over2})},
\end{equation}
where we have used the Euler representation for the gamma function.
We next carry out the Gaussian integration over $k$:
\begin{equation}
u=-{1\over4\sqrt{\pi}}{1\over(4\pi)^{d/2}}\sum_n\int_0^\infty{dt\over t}
t^{-1/2-d/2}e^{-t n^2\pi^2/a^2}.\label{aftergauss}
\end{equation}
Finally, we again use the Euler representation, and carry out the sum over
$n$ by use of the definition of the Riemann zeta function:
\begin{equation}
u=-{1\over4\sqrt{\pi}}{1\over(4\pi)^{d/2}}\left({\pi\over a}\right)^{d+1}
\Gamma\left(-{d+1\over2}\right)\zeta(-d-1).
\label{zeroptresult}
\end{equation}
When $d$ is an odd integer, this expression is indeterminate, but we can use
the reflection property
\begin{equation}
\Gamma\left({z\over2}\right)\zeta(z)\pi^{-z/2}=\Gamma\left({1-z\over2}\right)
\zeta(1-z)\pi^{(z-1)/2}
\label{refl}
\end{equation}
to rewrite (\ref{zeroptresult}) as
\begin{equation}
u=-{1\over2^{d+2}\pi^{d/2+1}}{1\over a^{d+1}}\Gamma\left(1+{d\over2}\right)
\zeta(2+d).\label{scalarenergy}
\end{equation}
We note that analytic continuation in $d$ is involved here: 
Eq.~(\ref{aftergauss}) is only valid if $\mbox{Re}\, d < -1$
 and the subsequent definition of the zeta function
is only valid if $\mbox{Re}\, d < -2$.  In the physical applications, $d$ is a 
positive integer.

We evaluate this general result (\ref{scalarenergy})
 at $d=2$.  This gives for the energy
per unit area in the transverse direction
\begin{equation}
u=-{\pi^2\over1440}{1\over a^3},
\label{casimir}
\end{equation}
where we have recalled that $\zeta(4)={\pi^4/90}$.
The force per unit area between the plates is
obtained by taking the negative derivative of $u$ with respect to $a$:
\begin{equation}
f_s=-{\partial\over\partial a}u=-{\pi^2\over 480}{1\over a^4}.
\label{casforce}
\end{equation}

The above result (\ref{casforce}) represents the Casimir force due to
a scalar field.  It is tempting  (and, in this case, is correct)
to suppose that to obtain the force due to electromagnetic field 
fluctuations between
parallel conducting plates, we simply multiply by a factor of 2 to account
for the two polarization states of the photon.  Doing so reproduces the
classic result of Casimir \cite{Cas}:
\begin{equation}
f_{\rm em}=-{\pi^2\over 240}{1\over a^4}.
\label{casimirclassic}
\end{equation}
A rigorous derivation of this result will be given in Sec.~\ref{sec:em}.
 
 \subsection{Scalar Green's Function}
\label{scalargreen}

We now rederive the result of Sec.~\ref{dimreg} by a physical and rigorous
Green's function approach.  The equation of motion of a massless
scalar field $\phi$ produced by a source $K$ is
\begin{equation}
-\partial^2\phi=K,
\label{eom}
\end{equation}
from which we deduce the equation satisfied by the corresponding
Green's function
\begin{equation}
-\partial^2 G(x,x')=\delta(x-x').
\label{kg}
\end{equation}
For the geometry shown in Fig.~\ref{fig1}, we introduce a reduced
Green's function $g(z,z')$ according to the Fourier transformation
\begin{equation}
G(x,x')=\int{d^d k\over(2\pi)^d}e^{i{\bf k}\cdot({\bf x-x'})}
\int{d\omega\over2\pi}e^{-i\omega(t-t')} g(z,z'),
\end{equation}
where we have suppressed the dependence of $g$ on ${\bf k}$ and $\omega$,
and have allowed $z$ on the right hand side
 to represent the coordinate perpendicular to the plates.
The reduced Green's function satisfies
\begin{equation}
\left(-{\partial^2\over\partial z^2}-\lambda^2\right)g(z,z')=\delta(z-z'),
\label{scalargreeneq}
\end{equation}
where $\lambda^2=\omega^2-k^2$.  Equation (\ref{scalargreeneq}) is to be
solved subject to the boundary conditions (\ref{eq:bc}), or
\begin{equation}
g(0,z')=g(a,z')=0.
\label{greenbc}
\end{equation}

We solve (\ref{scalargreeneq}) by the standard discontinuity method.
The form of the solution is
\begin{equation}
g(z,z')=\left\{\begin{array}{ll}
A\sin\lambda z,&0<z<z'<a,\\
B\sin\lambda(z-a),&a>z>z'>0,
\end{array}\right.
\end{equation}
which makes use of the boundary condition (\ref{greenbc}) on the plates.  
According to Eq.~(\ref{scalargreeneq}), $g$ is continuous
at $z=z'$, but its derivative has a discontinuity:
\alpheqn
\begin{eqnarray}
A\sin\lambda z'-B\sin\lambda(z'-a)&=&0,\label{cont}\\
A\lambda\cos\lambda z'-B\lambda\cos\lambda(z'-a)&=&1.\label{discont}
\end{eqnarray}
\reseteqn
The solution to this system of equations is
\alpheqn
\begin{eqnarray}
A&=&-{1\over\lambda}{\sin\lambda(z'-a)\over\sin\lambda a},\\
B&=&-{1\over\lambda}{\sin\lambda z'\over\sin\lambda a},
\end{eqnarray}
\reseteqn
which implies that the reduced Green's function is
\begin{equation}
g(z,z')=-{1\over\lambda\sin\lambda a}\sin\lambda z_<\sin\lambda (z_>-a),
\label{scgreen}
\end{equation}
where $z_>$ ($z_<$) is the greater (lesser) of $z$ and $z'$.

From knowledge of the Green's function we can calculate the force on the
bounding surfaces from the energy-momentum or stress tensor.  For a
scalar field, the stress tensor\footnote{The ambiguity in defining the
stress tensor has no effect.  We can add to $T_{\mu\nu}$ an arbitrary
multiple of $(\partial_\mu\partial_\nu-g_{\mu\nu}\partial^2)\phi^2$ 
\cite{schwingerpsf,ccj}.  But the $zz$ component of this tensor on
the surface vanishes by virtue of Eq.~(\ref{eq:bc}).} 
 is given by Eq.~(\ref{stresstensor}).
%\begin{equation}
%t_{\mu\nu}=\partial_\mu\phi\partial_\nu\phi+g_{\mu\nu}{\cal L},\label{st}
%\end{equation}
%where the Lagrange density is
%\begin{equation}
%{\cal L}=-{1\over2}\partial_\lambda\phi \partial^\lambda\phi.
%\end{equation}
What we require is the vacuum expectation value of $T_{\mu\nu}$
which can be obtained from the Green's function according
to Eq.~(\ref{vevgreen}), or
\begin{equation}
\langle\phi(x)\phi(x')\rangle={1\over i}G(x,x'),
\end{equation}
a time-ordered product being understood in the vacuum expectation value.
By virtue of the boundary condition (\ref{eq:bc}) we
compute  the normal-normal component of the
Fourier transform of the
stress tensor on the boundaries to be
\begin{equation}
\langle T_{zz}\rangle={1\over2i}\partial_z\partial_{z'}g(z,z')
\bigg|_{z\to z'=0,a}={i\over2}\lambda\cot\lambda a.\label{txx}
\end{equation}
We now must integrate on the transverse momentum and the frequency to get the
force per unit area.  The latter integral is best done by performing
a complex frequency rotation,
\begin{equation}
\omega\to i\zeta,\quad \lambda\to i\sqrt{k^2+\zeta^2}\equiv i\rho.
\end{equation}
Thus, the force per unit area is given by
\begin{equation}
f=-{1\over2}\int{d^d k\over(2\pi)^d}\int{d\zeta\over2\pi}\rho\coth\rho a.
\label{xscalarint}
\end{equation}
This integral does not exist. 

\begin{figure}
\centerline{
\psfig{figure=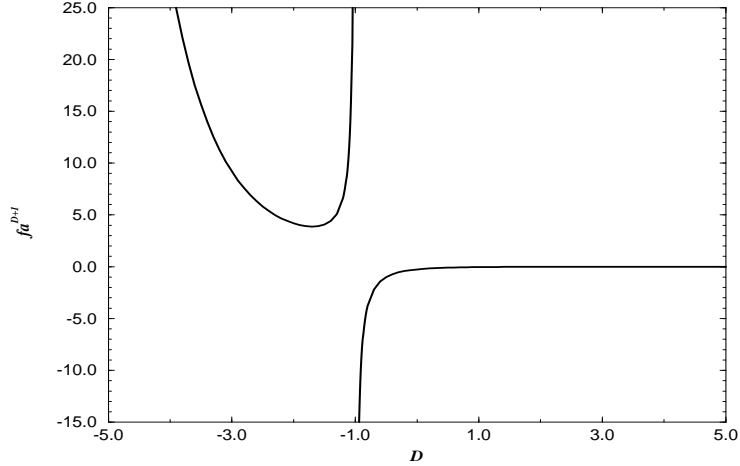,height=4.5in,width=3in,angle=270}}
\caption{A plot of the Casimir force per unit area $f$ in
Eq.~(\protect\ref{scalarforce})
for $-5<D<5$ for the case of a slab geometry (two parallel plates).
Here $D=d+1$.}
\label{fig2}
\end{figure}

What we do now is regard the right boundary at $z=a$, for example,
to be a perfect conductor of infinitesimal thickness, 
and consider the flux of momentum to the
right of that surface.  To do this, we find the Green's function which
vanishes at $z=a$, and has outgoing boundary conditions as $z\to\infty$,
$\sim e^{ikz}$.  A calculation just like that which led to Eq.~(\ref{scgreen})
yields for $z$, $z'>a$,
\begin{equation}
g(z,z')={1\over\lambda}\left[\sin\lambda(z_<-a) e^{i\lambda(z_>-a)}\right].
\label{scgreenout}
\end{equation}
(This obviously has the correct discontinuity $-{\partial \over\partial z}
g|_{z'-0}^{z'+0}=1$ and satisfies the appropriate boundary conditions.)
The corresponding normal-normal component of the stress tensor at $z=a$ is
\begin{equation}
\langle T_{zz}\rangle\bigg|_{z=z'=a}={1\over2i}\partial_z\partial_{z'}
g(z,z')\bigg|_{z=z'=a}={\lambda\over 2}.
\label{txxout}
\end{equation}
So, from the discontinuity in $T_{zz}$, that is, the difference
between (\ref{txx}) and (\ref{txxout}), we find the force per unit area
on the conducting surface:
\begin{equation}
f=-{1\over2}\int{d^d k\over(2\pi)^d}\int{d\zeta\over2\pi}\rho(\coth\rho a-1).
\label{0tempforce}
\end{equation}
 We evaluate this integral using polar
coordinates:
\begin{equation}
f=-{A_{d+1}\over(2\pi)^{d+1}}\int_0^\infty\rho^d\,d\rho{\rho\over e^{2\rho a}
-1}.\label{scalarint}
\end{equation}
Here $A_n$ is the surface area of a unit sphere in $n$ dimensions, which is
most easily found by integrating the multiple Gaussian integral
\begin{equation}
\int_{-\infty}^\infty d^nx \,e^{-x^2}=\pi^{n/2}
\end{equation}
in polar coordinates.  The result is
\begin{equation}
A_n={2\pi^{n/2}\over\Gamma(n/2)}.
\label{an}
\end{equation}
When we substitute this into (\ref{scalarint}), employ the integral
\begin{equation}
\int_0^\infty dy {y^{s-1}\over e^y-1}=\Gamma(s)\zeta(s),
\label{gammazeta}
\end{equation}
and use the identity
\begin{equation}
\Gamma(2z)=(2\pi)^{-1/2}2^{2z-1/2}\Gamma\left(z\right)\Gamma\left(
z+{1\over2}\right),
\label{dupform}
\end{equation}
we find for the force per unit transverse area
\begin{equation}
f=-(d+1)2^{-d-2}\pi^{-d/2-1}{\Gamma\left(1+{d\over2}\right)\zeta(d+2)\over
a^{d+2}}.\label{scalarforce}
\end{equation}
Evidently, Eq.~(\ref{scalarforce}) is the negative derivative of the Casimir
energy (\ref{scalarenergy}) with respect to the separation between the plates:
\begin{equation}
f=-{\partial u\over\partial a};
\end{equation}
this result has now been obtained by a completely well-defined approach.
The force per unit area, Eq.~(\ref{scalarforce}), 
is plotted in Fig.~\ref{fig2},
where $a\to2a$ and $d=D-1$.
This result was first derived by Ambj\o rn and Wolfram \cite{ambjorn}.

We can also derive the same result by computing the energy from the
energy-momentum tensor.\footnote{Again, the ambiguity in the
stress tensor is without effect, because the extra term here is 
$\nabla^2\phi^2$, which upon integration over space 
becomes a vanishing surface integral.}
  The relevant component is\footnote{As noted after Eq.~(\ref{stresstensor}),
we would get the same integrated energy if we dropped the second, Lagrangian,
term in $T_{00}$ there, that is, used $T_{00}=\partial_0\phi\partial_0\phi$.}
\begin{equation}
T_{00}={1\over2}(\partial_0\phi\partial_0\phi+\partial_1\phi\partial_1\phi+
\partial_2\phi\partial_2\phi+\partial_3\phi\partial_3\phi),
\end{equation}
so when the vacuum expectation value is taken, we find from Eq.~(\ref{scgreen})
\begin{eqnarray}
\langle T_{00}\rangle&=&-{1\over2i\lambda}{1\over\sin\lambda a}
[(\omega^2+k^2)\sin\lambda z\sin\lambda(z-a)\nonumber\\
&&\quad\quad\qquad\mbox{}+\lambda^2\cos\lambda z
\cos\lambda(z-a)]\nonumber\\
&=&-{1\over2i\lambda\sin\lambda a}[\omega^2\cos\lambda a-k^2\cos\lambda(2z-a)].
\label{t00form}
\end{eqnarray}
We now must integrate this over $z$ to find the energy per area between
the plates.  Integration of the second term in Eq.~(\ref{t00form}) gives
a constant, independent of $a$, which will not contribute to the force.
 The first term gives
\begin{equation}
\int_0^a dx\,\langle T_{00}\rangle=-{\omega^2 a\over 2 i \lambda}\cot\lambda a.
\end{equation}
As above, we now integrate over $\omega$ and $k$, after we perform the
complex frequency rotation.  We obtain
\begin{equation}
u=-{a\over2}\int{d^d k\over(2\pi)^d}\int{d\zeta\over 2\pi}{\zeta^2\over\rho}
\coth\rho a.
\end{equation}
If we introduce polar coordinates so that $\zeta=\rho\cos\theta$, we see
that this differs from Eq.~(\ref{xscalarint}) by the factor of $a
\langle\cos^2\theta\rangle$.  Here
\begin{equation}
\langle\cos^2\theta\rangle=
{\int_0^\pi\cos^2\theta\sin^{d-1}\theta\,d\theta
\over\int_0^\pi\sin^{d-1}\theta\,d\theta}={1\over d+1},
\end{equation}
which uses the integral
\begin{equation}
\int_0^\pi\sin^{d-1}\theta\,d\theta=\int_{-1}^1(1-x^2)^{(d-2)/2}dx=
2^{d-1}{\Gamma\left({d\over2}\right)^2\over\Gamma(d)}.
\end{equation}
Thus, we again recover Eq.~(\ref{scalarenergy}).

For the sake of completeness, we note that it is also possible to
use the eigenfunction expansion for the reduced Green's function.
That expansion is
\begin{equation}
g(z,z')={2\over a}\sum_{n=1}^\infty{\sin (n\pi z/a) \sin (n\pi z'/a)\over
n^2\pi^2/a^2-\lambda^2}.
\end{equation}
When we insert this into the stress tensor we encounter
\begin{equation}
\partial_z\partial_{z'}g(z,z')|_{z=z'=0,a}
={2\over a}\sum_{n=1}^\infty{n^2\pi^2/a^2\over n^2\pi^2/a^2-\lambda^2}.
\end{equation}
We subtract and add $\lambda^2$ to the numerator of this divergent sum,
and omit the divergent part, which
is simply a constant in $\rho$. Such terms correspond to $\delta$ functions
in space and time (contact terms), 
and should be omitted, since we are considering the
{\it limit\/} as the space-time points coincide.
We evaluate the resulting finite sum by use of the following expression
for the cotangent:
\begin{equation}
\cot\pi x={1\over\pi x}+{2x\over\pi}\sum_{k=1}^\infty{1\over x^2-k^2}.
\label{cotan}
\end{equation}
So in place of Eq.~(\ref{txx}) we obtain
\begin{equation}
\langle T_{zz}\rangle={i\over2}\lambda\left(\cot\lambda a-{1\over\lambda a}
\right),
\end{equation}
which agrees with Eq.~(\ref{txx}) apart from  an additional contact term.

\subsection{Massive Scalar}
\label{mass}
\begin{figure}
\centerline{
\psfig{figure=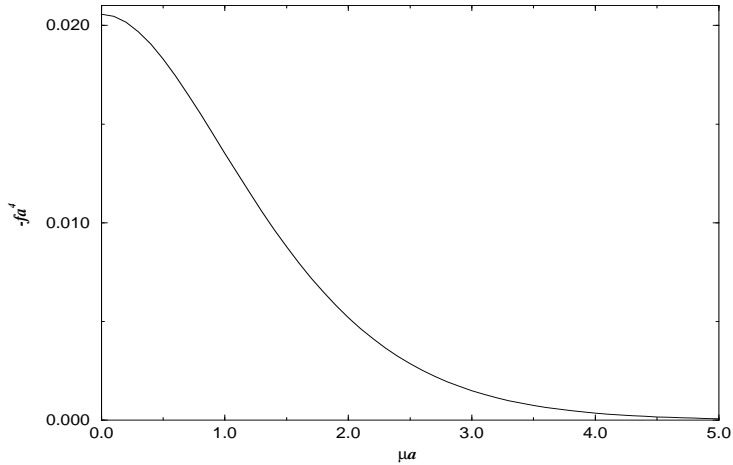,height=4.5in,width=3in,angle=270}}
\caption{Scalar Casimir force between parallel plates as a function of mass
for $d=2$.}
% Here $f=-{1\over32\pi^2}\int_{2\mu a}^\infty dx\, x^2
%\sqrt{x^2-4\mu^2 a^2}(e^x-1)^{-1}$.}
\label{fig:mass}
\end{figure}

It is easy to modify the discussion of Sec.~\ref{scalargreen}
to include a mass $\mu$ for the scalar field.  The reduced Green's function
now satisfies the equation
\begin{equation}
\left(-{\partial^2\over\partial z^2}-\lambda^2+\mu^2\right)g(z,z')=\delta
(z-z'),
\end{equation}
instead of Eq.~(\ref{scalargreeneq}), so the reduced Green's function between 
the plates is
\begin{equation}
g(z,z')=-{1\over\kappa}{\sin\kappa z_<\sin\kappa(z_>-a)\over\sin\kappa a}
\end{equation}
where
\begin{equation}
\kappa^2=\lambda^2-\mu^2.
\end{equation}
The calculation proceeds just as in Sec.~\ref{scalargreen}, and we find,
in place of Eq.~(\ref{scalarint})
\begin{equation}
f=-{A_{d+1}\over(2\pi)^{d+1}}\int_0^\infty\rho^d\,d\rho{\sqrt{\rho^2+\mu^2}
\over e^{2a\sqrt{\rho^2+\mu^2}}-1}.
\end{equation}
When we substitute the value of $A_{d+1}$ given by 
Eq.~(\ref{an}), and introduce
a dimensionless integration variable, we find for the force per unit area
\begin{equation}
f=-{1\over 2^{2(d+1)}\pi^{(d+1)/2}\Gamma\left({d+1\over2}\right)a^{d+2}}
\int_{2\mu a}^\infty dx\, x^2 {(x^2-4\mu^2 a^2)^{(d-1)/2}\over
e^x-1}.
\end{equation}
For $d=2$ this function is plotted in Fig.~\ref{fig:mass}.
The massive scalar was first treated by Hays in two dimensions \cite{hays},
and again done in general number of dimensions
by Ambj\o rn and Wolfram \cite{ambjorn}.

\subsection{Finite Temperature}
\label{sec:temp}
We next turn to a consideration of the Casimir effect at nonzero temperature.
In this case, fluctuations arise not only from quantum mechanics but from
thermal effects.  In fact, as we will shortly see, the high-temperature
limit is a purely classical phenomenon.

Formally, we can easily obtain the expression for the Casimir force 
between parallel plates at nonzero temperature.  In Eq.~(\ref{0tempforce})
we replace the imaginary frequency by
\begin{equation}
\zeta\to\zeta_n={2\pi n\over\beta},
\end{equation}
where $\beta=1/kT$, $T$ being the absolute temperature.  Correspondingly,
the frequency integral is replaced by a sum on the integer $n$:
\begin{equation}
\int{d\zeta\over2\pi}={1\over\beta}\sum_{n=-\infty}^\infty.
\end{equation}
Thus, Eq.~(\ref{0tempforce}) is replaced by
\begin{equation}
f^T=-{1\over2\beta}\int{d^dk\over(2\pi)^d}\sum_{n=-\infty}^\infty
{2\rho_n\over e^{2\rho_n a} -1},
\label{generaltemp}
\end{equation}
where $\rho_n=\sqrt{k^2+(2\pi n/\beta)^2}$.

We first consider the high-temperature limit.  When $T\to\infty$ ($\beta\to0$),
apart from exponentially small corrections, the contribution comes from the
$n=0$ term in the sum in Eq.~(\ref{generaltemp}).  That integral is easily
worked out in polar coordinates using Eqs.~(\ref{an}) and (\ref{gammazeta}).
The result is 
\begin{equation}
f^{T\to\infty}\sim-kT{d\over(2\sqrt{\pi}a)^{d+1}}
\Gamma\left({d+1\over2}\right)
\zeta(d+1).
\label{hitemp}
\end{equation}
In particular, for two and three dimensions, $d=1$ and $d=2$, respectively,
\alpheqn
\begin{eqnarray}
d=1:\quad f^{T\to\infty}&\sim&-kT{\pi\over24a^2},\\
d=2:\quad f^{T\to\infty}&\sim&-kT{\zeta(3)\over8\pi a^3}.
\end{eqnarray}
\reseteqn
This high-temperature limit should be classical.  Indeed, we can derive this
same result from the classical 
limit of statistical mechanics.  The Helmholtz free energy
for massless bosons is
\begin{equation}
F=-kT\ln Z,\quad \ln Z=-\sum_i\ln(1-e^{-\beta p_i}),
\end{equation}
from which the pressure on the plates can be obtained by differentiation:
\begin{equation}
p=-{\partial F\over\partial V}.
\end{equation}
We make the momentum-space sum explicit for our $d+1$ spatial geometry:
\begin{equation}
F=kTV\int{d^dk\over(2\pi)^{d+1}}{\pi\over a}\sum_{n=-\infty}^\infty
\ln(1-\exp(-\beta\sqrt{k^2+n^2\pi^2/a^2})).
\end{equation}
Now, for high temperature, $\beta\to0$, we expand the exponential, and
keep the first order term in $\beta$. We can write the result as
\begin{equation}
F=VkT{1\over2a}{d\over ds}\int{d^dk\over(2\pi)^d}\sum_{n=-\infty}^\infty
{1\over2}\beta^{2s}\left({n^2\pi^2\over a^2}+k^2\right)^s\bigg|_{s=0},
\end{equation}
where we have used the identity
\begin{equation}
\ln\xi={d\over ds}\xi^s\bigg|_{s=0}.
\end{equation} 
This trick allows us to proceed as in Sec.~\ref{dimreg}.
After the $k$ integration is done,
the $s$ derivative acts only on $1/\Gamma(-s)$:
\begin{equation}
{d\over ds}{1\over\Gamma(-s)}\bigg|_{s=0}=-1,
\end{equation}
so we easily find the result from Eq.~(\ref{refl})
\begin{equation}
F=-V{kT\over(2a\sqrt{\pi})^{d+1}}\zeta(d+1)\Gamma\left({1+d\over2}\right).
\end{equation}
The pressure, the force per unit area on the plates, is obtained by
applying the following differential operator to the free energy:
\begin{equation}
-{\partial\over\partial V}=-{1\over A}{\partial\over\partial a},
\end{equation}
where $V=Aa$, $A$ being the $d$-dimensional area of the plates.
The result of this operation coincides with Eq.~(\ref{hitemp}).

The low-temperature limit ($T\to0$ or $\beta\to\infty$) is more complicated
because $f^T$ is not analytic at $T=0$.  The most convenient way to proceed
is to resum the series in Eq.~(\ref{generaltemp}) by means of the Poisson sum
formula, which says that if $c(\alpha)$ is the Fourier transform of $b(x)$,
\begin{equation}
c(\alpha)={1\over2\pi}\int_{-\infty}^\infty b(x) e^{-i\alpha x}\, dx,
\label{fourier}
\end{equation}
then the following identity holds:
\begin{equation}
\sum_{n=-\infty}^\infty b(n)=2\pi\sum_{n=-\infty}^\infty c(2\pi n).
\label{poisson}
\end{equation}
Here, we take
\begin{equation}
b(n)=\int{d^d k\over(2\pi)^d}{2\rho_n\over e^{2\rho_n a}-1}.
\end{equation}
Introducing polar coordinates for $\bf k$, changing from $k$ 
to the dimensionless integration variable $z=2a\rho_x$,
and interchanging the order of $x$ and $z$ integration, we find for the
Fourier transform
\begin{equation}
c(\alpha)={A_d\over2^{2d}\pi^{d+1}a^{d+1}}\int_0^\infty{dz\,z^2\over e^z-1}
\int_0^{\beta z/4\pi a}\!\!\!\!\!\!\!\!dx\,\cos\alpha x \left(z^2-\left({4\pi
 a x\over\beta}\right)^2\right)^{(d-2)/2}.
\label{lowtemp1}
\end{equation}
The $x$ integral in Eq.~(\ref{lowtemp1}) is easily expressed in terms of a
Bessel function: It is \cite{grad}
\begin{eqnarray}
&&{\beta z^{d-1}\over4\pi a}\int_0^1du\,\cos\left({\alpha\beta z\over4\pi a}u
\right)(1-u^2)^{(d-2)/2}\nonumber\\
=&&\alpha^{(1-d)/2}z^{(d-1)/2}\left({8\pi a\over\beta}\right)^{(d-3)/2}
\!\!\!
\sqrt\pi\,\Gamma\left({d\over2}\right)J_{(d-1)/2}\left({\alpha\beta z\over
4\pi a}\right).
\label{lowtemp2}
\end{eqnarray}
We thus encounter the $z$ integral
\begin{equation}
I(s)=\int_0^\infty{dz\,z^{(d+3)/2}\over e^z-1}J_{(d-1)/2}(sz),
\label{iofs}
\end{equation}
where $s=\alpha\beta /4\pi a$.

The zero-temperature limit comes entirely from $\alpha=0$:
\begin{equation}
f^{T=0}=-{\pi\over\beta}c(0).
\end{equation}
So we require the small-$x$ behavior of the Bessel function,
\begin{equation}
J_{(d-1)/2}(x)\sim\left({x\over2}\right)^{(d-1)/2}{1\over\Gamma\left(
{d+1\over2}\right)},\quad x\to0,
\end{equation}
whence using Eq.~(\ref{gammazeta})
\begin{equation}
I(s)\sim\left({s\over2}\right)^{(d-1)/2}{2^{d+1}\over\sqrt\pi}
{d+1\over2}\Gamma\left({d\over2}+1\right)\zeta(d+2), \quad s\to0.
\end{equation}
Inserting this into Eq.~(\ref{lowtemp1}) 
we immediately recover the zero-temperature result (\ref{scalarforce}).

We now seek the leading correction to this.  We rewrite $I(s)$ as
\begin{eqnarray}
I(s)&=&{1\over s^{(d+5)/2}}\int_0^\infty{dy\,y^{(d+3)/2}\over1-e^{-y/s}}e^{-y/s}
J_{(d-1)/2}(y)\nonumber\\
&=&{1\over s^{(d+5)/2}}\int_0^\infty dy\,y^{(d+3)/2}e^{-y/s}\sum_{n=0}^\infty
e^{-ny/s}J_{(d-1)/2}(y),
\end{eqnarray}
where we have employed the geometric series.  The Bessel-function integral
has an elementary form \cite{grad}:
\begin{eqnarray}
\int_0^\infty dy\,J_q(y) e^{-by}={(\sqrt{b^2+1}-b)^q\over\sqrt{b^2+1}}
\end{eqnarray} 
is the fundamental integral, and the form we want can be written as
\begin{equation}
\int_0^\infty dy\,J_q(y) y^p e^{-by}=
(-1)^p\left(d\over db\right)^p{(\sqrt{b^2+1}-b)^q\over\sqrt{b^2+1}},
\end{equation} 
provided $p$ is a nonnegative
 integer.  (For the application here, this means $d$ is
odd, but we will be able to analytically continue the final result to
arbitrary $d$.)
Then we can write $I(s)$ in terms of the series
\begin{equation}
I(s)={(-1)^{(d+3)/2}\over s^{(d+5)/2}}\sum_{l=1}^\infty F(l), \quad l=n+1,
\label{iofssum}
\end{equation}
where 
\begin{equation}
F(l)=\left(d\over d l/s\right)^{
(d+3)/2}{\left[\sqrt{\left(l\over s \right)^2+1}
-{l\over s}\right]^{(d-1)/2}\over\sqrt{\left(l\over s \right)^2+1}}.
\label{fol}
\end{equation}
We evaluate the sum in Eq.~(\ref{iofssum}) by means of the Euler-Maclaurin 
summation formula, which has the following formal expression:
\begin{eqnarray}
\sum_{l=1}^\infty F(l)&=&\int_1^\infty dl\,F(l)+{1\over2}[F(\infty)+F(1)]
\nonumber\\
&&\mbox{}
+\sum_{k=1}^\infty{1\over(2k)!}B_{2k}[F^{(2k-1)}(\infty)-F^{(2k-1)}(1)].
\label{eulermac}
\end{eqnarray}
Here, $B_n$ represents the $n$th Bernoulli number.
Since we are considering $\alpha\ne0$ (the $\alpha=0$ term was dealt with in
the previous paragraph), the low-temperature limit corresponds to the limit
$s\to\infty$.  It is easy then to see that $F(\infty)$ and all its derivatives
there vanish. The function $F$ at 1 has the general form
\begin{equation}
F(1)=\left(d\over d\epsilon\right)^{(d+3)/2}{\left[\sqrt{\epsilon^2+1}
-\epsilon\right]^{(d-1)/2}\over\sqrt{\epsilon^2+1}}, \quad\epsilon=1/s\to0.
\end{equation}
 By examining the various possibilities for odd $d$, $d=1$,
$d=3$, $d=5$, and so on, we  find the result
\begin{equation}
F(1)=-(-1)^{(d-1)/2}d!!=-(-1)^{(d-1)/2}2^{(d+1)/2}\pi^{-1/2}\Gamma
\left({d\over2}+1\right).
\end{equation}
Because it is easily seen that before the limit $\epsilon\to0$ is taken,
$F(1)$ is an even function of $\epsilon$, it follows that the odd derivatives
of $F$ evaluated at 1 that appear in Eq.~(\ref{eulermac}) vanish.
Finally, the integral in Eq.~(\ref{eulermac}) is that of a total derivative:
\begin{equation}
\int_1^\infty dl\,F(l)=-{1\over\epsilon}\left(d\over d\epsilon\right)^{(d+1)/2}
{\left[\sqrt{\epsilon^2+1}
-\epsilon\right]^{(d-1)/2}\over\sqrt{\epsilon^2+1}}
=-F(1).
\end{equation}
Thus, the final expression for $I(s)$ is
\begin{equation}
I(s)\sim{1\over s^{(d+5)/2}} 2^{(d-1)/2}\pi^{-1/2}\Gamma\left({d\over2}
+1\right).
\end{equation}
Note that a choice of analytic continuation to $d$ other than an odd integer 
has been made so as to avoid oscillatory behavior in $d$.

We return to Eq.~(\ref{lowtemp1}).  It may be written as
\begin{eqnarray}
c(\alpha)&=&{2\pi^{d/2}\over\Gamma(d/2)}{1\over 2^{2d}\pi^{d+1}a^{d+1}}
\alpha^{(1-d)/2}
\left(8\pi a\over\beta\right)^{(d-3)/2}\sqrt\pi\,\Gamma(d/2)I\left(\alpha
\beta\over4\pi a\right)\nonumber\\
 &=&2^{-d-1}\pi^{-d/2-1}\left(4\pi\over\beta\right)^{d+1}
\Gamma\left({d\over2}+1\right){1\over \alpha^{d+2}}.
\end{eqnarray}
The correction to the zero-temperature result Eq.~(\ref{scalarforce})
is obtained from
\begin{eqnarray}
f^{T\to0}_{\rm corr}&=&-{2\pi\over\beta}\sum_{n=1}^\infty c(2\pi n)
\nonumber\\
&=&-\pi^{-d/2-1}\Gamma(d/2+1)\zeta(d+2)\beta^{-d-2}.
\end{eqnarray}
Thus, the force per unit area in the low-temperature limit has the
form
\begin{equation}
f^{T\to0}\approx
-{(d+1)2^{-d-2}\pi^{-d/2-1}\over a^{d+2}}\Gamma(d/2+1)\zeta(d+2)
\left[1+{1\over d+1}\left(2 a\over\beta\right)^{d+2}\right],
\end{equation}
of which the $d=1$ and $d=2$ cases are familiar:
\alpheqn
\begin{eqnarray}
d=1: &\quad&f^{T\to0}\approx
 -{1\over8\pi a^3}\zeta(3)\left[1+4{a^3\over\beta^3}\right],\\
d=2: &\quad&f^{T\to0}\approx
 -{\pi^2\over480 a^4}\left[1+{16\over3}{a^4\over\beta^4}\right].
\end{eqnarray}
\reseteqn
These equations are incomplete in that they omit exponentially small
terms; for example, in the last square bracket, we should add
the term
\begin{equation}
-{240\over\pi}{a\over\beta}e^{-\pi\beta/a}.
\end{equation}

Finite temperatures were first discussed by Lifshitz \cite{lifshitz}, 
but then considered more fully by Fierz, Sauer, and Mehra 
\cite{fierz,sauer,mehra}. Hargreaves \cite{hargreaves} analyzed the
discrepancy between the results of Refs.~\cite{lifshitz} and \cite{sauer}.

\subsection{Electromagnetic Casimir Force}
\label{sec:em}
We now turn to the situation originally treated by Casimir:  the force
between parallel conducting plates due to quantum fluctuations in the 
electromagnetic field.  An elegant procedure, which can be applied to
much more complicated geometries, involves the introduction of the
Green's dyadic, defined as the response function between the (classical)
electromagnetic field and the polarization source (this formalism
is introduced in Refs.~\cite{sch1,sch2}):
\begin{equation}
{\bf E}(x)=\int (d{\bf x'})\,\bGamma(x,x')\cdot {\bf P}(x').
\label{egreen}
\end{equation}
In the following we will use the Fourier transform of $\bf\Gamma$ in
frequency:
\begin{equation}
\bGamma(x,x')=\int{d\omega\over2\pi}e^{-i\omega(t-t')}\bGamma({\bf r},
{\bf r}';\omega),
\end{equation}
which satisfies Maxwell's equations
\alpheqn
\begin{eqnarray}
\mbox{\boldmath$\nabla$}\times\bGamma&=&i\omega\bPhi,\label{max1}\\
-\bnabla\times\bPhi-i\omega\bGamma&
=&i\omega{\bf 1}\delta({\bf r}
-{\bf r}').
\label{max2}
\end{eqnarray}
\reseteqn
The second Green's dyadic appearing here is solenoidal,
\begin{equation}
\mbox{\boldmath$\nabla$}\cdot\bPhi=0,
\end{equation}
as is $\bGamma$ if a multiple of a $\delta$ function is subtracted:
\begin{equation}
\mbox{\boldmath$\nabla$}
\cdot\bGamma'=0,\quad \bGamma'=\bGamma+{\bf 1}\delta
({\bf r}-{\bf r}').
\label{sole}
\end{equation}
The system of first-order equations Eqs.~(\ref{max1}), (\ref{max2})
 can be easily converted
to second-order form:
\alpheqn
\begin{eqnarray}
(\nabla^2+\omega^2)\bGamma'&=&
-\mbox{\boldmath$\nabla$}\times(\mbox{\boldmath$\nabla$}\times
{\bf 1})\delta({\bf r}-{\bf r}'),\label{greendyads1}\\
(\nabla^2+\omega^2)\bPhi&=&i\omega\mbox{\boldmath$\nabla$}\times
{\bf 1}\delta({\bf r}-{\bf r}').
\label{greendyads2}
\end{eqnarray}
\reseteqn

The system of equations Eqs.~(\ref{greendyads1}), (\ref{greendyads2})
 is quite general.  We specialize
to the case of parallel plates by introducing the transverse Fourier
transform:
\begin{equation}
\bGamma'({\bf r}, {\bf r}';\omega)=\int{(d{\bf k})\over(2\pi)^2}
e^{i{\bf k}\cdot({\bf r}_\perp-{\bf r}'_\perp)}{\bf g}(z,z';{\bf k},\omega).
\label{ftgreen}
\end{equation}
The equations satisfied by the various Cartesian components of $\bf g$
may be easily worked out once it is recognized that
\begin{equation}
[\mbox{\boldmath$\nabla$}
\times(\mbox{\boldmath$\nabla$}\times{\bf 1})]_{ij}=\nabla_i\nabla_j
-\delta_{ij}\nabla^2.
\end{equation}
In terms of the Fourier transforms, these equations are
\alpheqn
\begin{eqnarray}
\left({\partial^2\over\partial z^2}-k^2+\omega^2\right)g_{zz}&=&
-k^2\delta(z-z'),\\
\left({\partial^2\over\partial z^2}-k^2+\omega^2\right)g_{zx}&=&
-ik_x{\partial\over\partial z}\delta(z-z'),\\
\left({\partial^2\over\partial z^2}-k^2+\omega^2\right)g_{zy}&=&
-ik_y{\partial\over\partial z}\delta(z-z'),\\
\left({\partial^2\over\partial z^2}-k^2+\omega^2\right)g_{xx}&=&
\left(-k_y^2+{\partial^2\over\partial z^2}\right)\delta(z-z'),\\
\left({\partial^2\over\partial z^2}-k^2+\omega^2\right)g_{yy}&=&
\left(-k_x^2+{\partial^2\over\partial z^2}\right)\delta(z-z'),\\
\left({\partial^2\over\partial z^2}-k^2+\omega^2\right)g_{xy}&=&
k_xk_y\delta(z-z').
\end{eqnarray}
\reseteqn

We solve these equations subject to the boundary condition that
the transverse components of the electric field vanish on the conducting
surfaces, that is,
\begin{equation}
{\bf n}\times{\bGamma}'|_{z=0,a}=0,
\end{equation}
where $\bf n$ is the normal to the surface.
That means any $x$ or $y$ components vanish at $z=0$ or at $z=a$.
Therefore, 
$g_{xy}$ is particularly simple.  By the standard discontinuity method,
we immediately find [cf.~Eq.~(\ref{scgreen})]
\begin{equation}
g_{xy}=g_{yx}={k_xk_y\over\lambda\sin\lambda a}(ss),
\end{equation}
where
\begin{equation}
(ss)=\sin\lambda z_<\sin\lambda(z_>-a).
\end{equation}
To find $g_{xx}$ we simply subtract a $\delta$ function:
\begin{equation}
g'_{xx}=g_{xx}-\delta(z-z').
\end{equation}
Then, we again find at once
\begin{equation}
g'_{xx}={k_x^2-\omega^2\over\lambda\sin\lambda a}(ss),
\end{equation}
and similarly
\begin{equation}
g'_{yy}={k_y^2-\omega^2\over\lambda\sin\lambda a}(ss).
\end{equation}
To determine the boundary condition on $g_{zz}$, we recall the solenoidal
condition on $\bGamma'$, Eq.~(\ref{sole}), which implies that
\begin{equation}
{\partial\over\partial z}g_{zz}\bigg|_{z=0,a}=0.
\label{gzzbc}
\end{equation}
This then leads straightforwardly to the conclusion
\begin{equation}
g_{zz}=-{k^2\over\lambda\sin\lambda a}(cc),
\end{equation}
where
\begin{equation}
(cc)=\cos\lambda z_<\cos\lambda(z_>-a).
\end{equation}
The remaining components have the property that the functions are
discontinuous, while, apart from a $\delta$ function, their derivatives
are continuous:
\begin{equation}
g_{zx}=-{ik_x\over\sin\lambda a}(cs),
\label{gzxcs}
\end{equation}
\begin{equation}
g_{zy}=-{ik_y\over\sin\lambda a}(cs),
\label{gzycs}
\end{equation}
where, because of the analogue of Eq.~(\ref{gzzbc}), or because the components
must vanish when $z'=0,a$,
\begin{equation}
(cs)=\left\{ \begin{array}{ll}
 \cos\lambda z\sin\lambda(z'-a),& z<z',\\
\sin\lambda z'\cos\lambda(z-a),&z>z'.
\end{array}\right.
\end{equation}
Similarly, 
\begin{equation}
g_{xz}={ik_x\over\sin\lambda a}(sc),
\label{gxzsc}
\end{equation}
\begin{equation}
g_{yz}={ik_y\over\sin\lambda a}(sc),
\label{gyzsc}
\end{equation}
where
\begin{equation}
(sc)=\left\{ \begin{array}{ll}
 \sin\lambda z\cos\lambda(z'-a), & z<z',\\
\cos\lambda z'\sin\lambda(z-a),& z>z'.
\end{array}\right.
\end{equation}
The relation between Eqs.~(\ref{gzxcs}), (\ref{gzycs}) and
Eqs.~(\ref{gxzsc}), (\ref{gyzsc}) just reflects the symmetry
\begin{equation}
{\bf \bGamma'(r,r')}=\bGamma^{\prime T}({\bf r',r}),
\end{equation}
where $T$ denotes transposition, which means that
\begin{equation}
g_{ij}(z,z';{\bf k})=g_{ji}(z',z;-{\bf k}).
\end{equation}

The normal-normal component of the electromagnetic stress tensor is
\begin{equation}
T_{zz}={1\over2}(E^2+H^2)-(E_z^2+H_z^2).
\end{equation}
The vacuum expectation value is obtained by the replacements
\alpheqn
\begin{equation}
\langle {\bf E}(x){\bf E}(x')\rangle={1\over i}{\bGamma}(x,x'),
\label{ereplace}
\end{equation}
\begin{equation}
\langle {\bf H}(x){\bf H}(x')\rangle=-{1\over i}{1\over\omega^2}
\mbox{\boldmath$\nabla$}\times\bGamma(x,x')\times
\overleftarrow{\mbox{\boldmath $\nabla$}}'.
\label{hreplace}
\end{equation}
\reseteqn
In terms of the Fourier transforms, we have
\begin{eqnarray}
\langle T_{zz}\rangle&=&{1\over2i\omega^2}\left[-(\omega^2-k^2)g_{zz}
+(\omega^2-k_y^2)g_{xx}+(\omega^2-k_x^2)g_{yy}\right.\nonumber\\
&&\mbox{}+ik_y\left({\partial\over\partial z}g_{yz}-{\partial\over\partial z'}
g_{zy}\right)
+ik_x\left({\partial\over\partial z}g_{xz}-{\partial\over\partial z'}g_{zx}
\right)\nonumber\\
&&\left.\mbox{}+k_xk_y(g_{xy}+g_{yx})+{\partial\over\partial z}{\partial\over
\partial z'}
(g_{xx}+g_{yy})\right].
\end{eqnarray}
When the appropriate Green's functions are inserted into the above,
enormous simplification occurs on the surface, and we are left with
\begin{equation}
\langle T_{zz}\rangle\big|_{z=z'=0,a}=i\lambda\cot\lambda a,
\end{equation}
which indeed is twice the scalar result (\ref{txx}), as claimed at the
end of Sec.~\ref{dimreg}. 

\subsection{Fermionic Casimir Force}
\label{fermion}
We conclude this section with a discussion of the force on parallel surfaces
due to fluctuations in a massless Dirac field.  For this simple
geometry, the primary distinction between this case and what has gone before
lies in the boundary conditions.  The boundary conditions appropriate to
the Dirac equation are the so-called bag-model boundary conditions.  That is,
if $n^\mu$ represents an outward normal at a boundary surface, the condition
on the Dirac field $\psi$ there is
\begin{equation}
(1+in\cdot \gamma)\psi=0.
\label{bagbc}
\end{equation}
For the situation of parallel plates at $z=0$ and $z=a$, this means
\begin{equation}
(1\mp i\gamma^3)\psi=0
\label{bcfermi}
\end{equation}
at $z=0$ and $z=a$, respectively.
In the following, we will choose a representation of the Dirac matrices
in which $i\gamma_5$ is diagonal, in $2\times 2$ block form, 
\begin{equation}
i\gamma_5=\left(\begin{array}{cc}
1&\,0\\
0&\,-1
\end{array}\right)
\end{equation}
while
\begin{equation}
\gamma^0=\left(\begin{array}{cc}
0&\,-i\\
i&\,0
\end{array}\right)
\end{equation}
from which the explicit form of all the other Dirac matrices follow
from
$\mbox{\boldmath $\gamma$ }=i\gamma^0\gamma_5\mbox{\boldmath$\sigma$}$.

\subsubsection{Summing modes}

It is easiest, but not rigorous, to sum modes as in Sec.~\ref{dimreg}.
We introduce a Fourier transform in time and the transverse spatial
directions,
\begin{equation}
\psi(x)=\int{d\omega\over2\pi}e^{-i\omega t}\int {(d{\bf k})\over(2\pi)^2}
e^{i{\bf k\cdot x}}\psi(z;{\bf k}, \omega),
\end{equation}
so that the Dirac equation for a massless fermion $-i\gamma\partial \psi=0$
becomes (if ${\bf k}$ lies along the negative $y$ axis)
\alpheqn
\begin{eqnarray}
\left(-\omega\mp i{\partial\over\partial z}\right)u_\pm \pm k v_\pm&=&0,
\label{deqn1}\\
\pm k u_\pm+\left(-\omega\pm i{\partial\over\partial z}\right)v_\pm&=&0.
\label{deqn2}
\end{eqnarray}
\reseteqn
where the subscripts indicate the eigenvalues of $i\gamma_5$
and $u$ and $v$ are eigenvectors of $\sigma^3$ with eigenvalue $+1$ or $-1$,
respectively.  This system of equations is to be solved to the boundary
conditions (\ref{bcfermi}), or
\alpheqn
\begin{eqnarray}
u_++u_-|_{z=0}&=&0,\\
v_+-v_-|_{z=0}&=&0,\\
u_+-u_-|_{z=a}&=&0,\\
v_++v_-|_{z=a}&=&0.
\end{eqnarray}
\reseteqn

The solution is straightforward.  Each component satisfies
\begin{equation}
\left({\partial^2\over\partial z^2}+\lambda^2\right)\psi=0,
\end{equation}
where $\lambda^2=\omega^2-k^2$, so that the components are expressed
as follows:
\alpheqn
\begin{eqnarray}
u_++u_-&=&A\sin\lambda z,\\
v_+-v_-&=&B\sin\lambda z,\\
u_+-u_-&=&C\sin\lambda(z-a),\\
v_++v_-&=&D\sin\lambda(z-a).
\end{eqnarray}
\reseteqn
Inserting these into the Dirac equation
(\ref{deqn1}) and (\ref{deqn2}), we find, first, an eigenvalue
condition on $\lambda$:
\begin{equation}
\cos\lambda a=0,
\end{equation}
or
\begin{equation}
\lambda a=(n+{1\over2})\pi,
\end{equation}
where $n$ is an integer. We then find 
 two independent solutions for the coefficients:
\alpheqn
\begin{eqnarray}
A&\ne&0,\\
B&=&0,\\
C&=&{i\omega \over\lambda}(-1)^n A,\\
D&=&{i k \over\lambda}(-1)^n A,
\end{eqnarray}
\reseteqn
and
\alpheqn
\begin{eqnarray}
A&=&0,\\
B&\ne&0,\\
C&=&{k \over i\lambda}(-1)^n B,\\
D&=&{\omega \over i\lambda}(-1)^n B.
\end{eqnarray}
\reseteqn
Thus, when we compute the zero-point energy, we must sum over
odd integers, noting that there are two modes, and  remembering the
characteristic minus sign for fermions:
Instead of Eq.~(\ref{zeropt2}),
the Casimir energy is
\begin{eqnarray}
u&=& -2{1\over2}\sum_{n=0}^\infty{(d{\bf k})\over(2\pi)^2}
\sqrt{k^2+{(n+1/2)^2\pi^2
\over a^2}}\nonumber\\
&=&{1\over2\sqrt{\pi}}{1\over4\pi}\sum_{n=0}^\infty\int_0^\infty {dt\over t}
t^{-3/2}e^{-t(n+1/2)^2\pi^2/a^2}\nonumber\\
&=&{1\over8\pi^{3/2}}\Gamma\left(-{3\over2}\right)\sum_{n=0}^\infty
{(n+1/2)^3\pi^3\over a^3}\nonumber\\
&=&-{\pi^2\over6 a^3}{7\over8}\zeta(-3),
\end{eqnarray}
which is ${7\over8}\times 2$ times the scalar result
(\ref{casimir}) because $\zeta(-3)=-B_4/4=1/120$.  (The factor of 2 refers to 
the two spin modes of the fermion.)

\subsubsection{Green's Function Method}

Again, a more controlled calculation starts from the equation
satisfied by the Dirac Green's function,
\begin{equation}
\gamma{1\over i}\partial G(x,x')=\delta(x-x'),
\end{equation}
subject to the boundary condition
\begin{equation}
(1+i{\bf n\cdot \mbox{\boldmath$\gamma$}})G\bigg|_{z=0,a}=0.
\end{equation}
We introduce a reduced, Fourier-transformed, Green's function,
\begin{equation}
G(x,x')=\int{d\omega\over2\pi}e^{-i\omega(t-t')}\int{(d{\bf k})\over(2\pi)^2}
e^{i{\bf k\cdot(x-x')}}g(z,z';{\bf k}, \omega),
\end{equation}
which satisfies
\begin{equation}
\left(-\gamma^0\omega+{\bf\mbox{\boldmath$\gamma$}\cdot k}+\gamma^3{1\over i}
{\partial\over
\partial z}\right)g(z,z')=\delta(z-z').
\end{equation}
Introducing the representation for the gamma matrices given above,
we find that the components of $g$ corresponding to the $+1$ or $-1$
eigenvalues of $i\gamma_5$,
\begin{equation}
g=\left(\begin{array}{cc}
g_{++}&g_{+-}\\
g_{-+}&g_{--}
\end{array}\right),
\end{equation}
satisfy the coupled set of equations
\alpheqn
\begin{eqnarray}
\left(-\omega\pm{\bf \mbox{\boldmath$\sigma$}
\cdot k}\mp i\sigma^3{\partial\over\partial z}
\right)g_{\pm\pm}&=&0,\\
 \left(-\omega\pm{\bf \mbox{\boldmath$\sigma$}
\cdot k}\mp i\sigma^3{\partial\over\partial z}
\right)g_{\pm\mp}&=&\mp i\delta(z-z').
\label{coupledfermi}
\end{eqnarray}
\reseteqn
We then resolve each of these components into eigenvectors of $\sigma^3$:
\begin{equation}
g_{\pm\pm}=\left(\begin{array}{cc}
u_{\pm\pm}^{(+)}&u_{\pm\pm}^{(-)}\\
v_{\pm\pm}^{(+)}&v_{\pm\pm}^{(-)}
\end{array}\right)
\end{equation}
and similarly for $g_{\pm\mp}$.  These components satisfy the coupled equations
\alpheqn
\begin{eqnarray}
\left(-\omega\mp i{\partial\over\partial z}\right)u_{\pm\pm}^{(\pm)}\pm k
 v_{\pm\pm}^{(\pm)}&=&0,\\
\pm k u_{\pm\pm}^{(\pm)}+\left(-\omega\pm i{\partial\over\partial z}\right)
v_{\pm\pm}^{(\pm)}&=&0,\\
\left(-\omega\mp i{\partial\over\partial z}\right)u_{\pm\mp}^{(+)}\pm k
 v_{\pm\mp}^{(+)}&=&\mp i\delta(z-z'),\\
\left(-\omega\mp i{\partial\over\partial z}\right)u_{\pm\mp}^{(-)}\pm k
 v_{\pm\mp}^{(-)}&=&0,\\
\pm k u_{\pm\mp}^{(+)}+\left(-\omega\pm i{\partial\over\partial z}\right)
v_{\pm\mp}^{(+)}&=&0,\\
\pm k u_{\pm\mp}^{(-)}+\left(-\omega\pm i{\partial\over\partial z}\right)
v_{\pm\mp}^{(-)}&=&\mp i\delta(z-z'),
\end{eqnarray}
\reseteqn
which aside from the inhomogeneous terms are replicas of (\ref{deqn1})
and (\ref{deqn2}).
These equations are to be solved subject to the boundary conditions
\alpheqn
\begin{eqnarray}
u_{++}^{(\pm)}-u_{-+}^{(\pm)}|_{z=a}&=&0,\\
u_{++}^{(\pm)}+u_{-+}^{(\pm)}|_{z=0}&=&0,\\
u_{+-}^{(\pm)}-u_{--}^{(\pm)}|_{z=a}&=&0,\\
u_{+-}^{(\pm)}+u_{--}^{(\pm)}|_{z=0}&=&0,\\
v_{++}^{(\pm)}+v_{-+}^{(\pm)}|_{z=a}&=&0,\\
v_{++}^{(\pm)}-v_{-+}^{(\pm)}|_{z=0}&=&0,\\
v_{+-}^{(\pm)}+v_{--}^{(\pm)}|_{z=a}&=&0,\\
v_{+-}^{(\pm)}-v_{--}^{(\pm)}|_{z=0}&=&0.
\end{eqnarray}
\reseteqn
Again, the solution is  straightforward.  We find
\alpheqn
\begin{eqnarray}
u_{++}^{(+)}&=&u_{--}^{(+)*}=v_{++}^{(-)*}=v_{--}^{(-)}\nonumber\\
&=&{1\over 2\cos\lambda a}\left[
\cos\lambda(z+z'-a)+{i\omega\over\lambda}\sin\lambda
(z+z'-a)\right],
\label{fermisolbegin}\\
v_{++}^{(+)}&=&v_{--}^{(+)}=u_{++}^{(-)*}=u_{--}^{(-)*}\nonumber\\
&=&{ik\over2\lambda}{1\over\cos\lambda a}\sin\lambda(z+z'-a),\\
%\end{eqnarray}
%\begin{eqnarray}
u_{-+}^{(+)}&=&u_{+-}^{(+)*}=-v_{-+}^{(-)*}=-v_{+-}^{(-)}\nonumber\\
&=&{1\over2\cos\lambda a}\left[\epsilon(z-z')\cos\lambda(z_>-z_<-a)
\right.\nonumber\\
&&\left.\mbox{}-{i\omega\over\lambda}\sin\lambda(z_>-z_<-a)\right],\\
v_{-+}^{(+)}&=&v_{+-}^{(+)}=u_{-+}^{(-)}=u_{+-}^{(-)}\nonumber\\
&=&{ik\over2\lambda}{1\over\cos\lambda a}\sin\lambda(z_>-z_<-a),
\label{fermisolend}
\end{eqnarray}
\reseteqn
where
\begin{equation}
\epsilon(z-z')=\left\{\begin{array}{ll}
1&\mbox{ if $z>z'$},\\
-1&\mbox{ if $z<z'$}.
\end{array}
\right.
\end{equation}

We now insert these Green's functions into the vacuum expectation value
of the energy-momentum tensor.  The latter is
\begin{equation}
T^{\mu\nu}={1\over2}\psi\gamma^0{1\over2}\left(\gamma^\mu{1\over i}\partial^\nu
+\gamma^\nu{1\over i}\partial^\mu\right)\psi+g^{\mu\nu}{\cal L}.
\end{equation}
We take the vacuum expectation value by the replacement
\begin{equation}
\psi\psi\gamma^0\to{1\over i}G,
\end{equation}
where $G$ is the fermionic Green's function computed above. 
Because we are interested in the {\it limit\/} as $x'\to x$ we can ignore
the Lagrangian term in the energy-momentum tensor, leaving us with
in the transform space
\def\Tr{\mbox{Tr }}
\def\tr{\mbox{tr }}
\begin{eqnarray}
\langle T^{33}\rangle&=&{1\over2}\Tr\gamma^3\partial_3 g(z,z')
\bigg|_{z'\to z}\nonumber\\
&=&{i\over2}{\partial\over\partial z}\tr\sigma^3(g_{-+}+g_{+-})\bigg|_{z'\to z}
\nonumber\\
&=&{i\over2}{\partial\over\partial z}[u_{-+}^{(+)}+u_{+-}^{(+)}-(v_{-+}^{(-)}
+v_{+-}^{(-)})]\bigg|_{z'\to z}.
\end{eqnarray}
When we insert the solution found above, 
Eqs.~(\ref{fermisolbegin})--(\ref{fermisolend}), 
we obtain
\begin{equation}
\langle T^{33}\rangle=2i{\partial\over\partial z}\mbox{Re}\, u_{-+}^{(+)}.
\end{equation}
Carrying out the differentiation and setting $z=z'$ we find
\begin{equation}
\langle T^{33}\rangle=i\lambda\tan\lambda a,
\end{equation}
where again we ignore the $\delta$-function term. [Compare the scalar
result (\ref{txx}).]

We now follow the same procedure given in Sec.~\ref{scalargreen}:
The force per unit area is
\begin{eqnarray}
f&=&\int{d^2k\over(2\pi)^2}\int{d\omega\over2\pi} i\lambda\tan\lambda a
\nonumber\\
&=&\int{d^2k\over(2\pi)^2}\int{d\zeta\over2\pi}\rho  \tanh\rho a\nonumber
\\
&=&{1\over 2\pi^2}\int_0^\infty d\rho\,\rho^3\left[1-{2\over e^{2\rho a}+1}
\right]
\end{eqnarray}
As in Eq.~(\ref{xscalarint}) we omit the $1$ in the last square bracket: 
The same term is present
in the vacuum energy outside the plates, so cancels out when we compute
the discontinuity across the plates.  We are left with, then,
\begin{equation}
f=-{1\over16\pi^2 a^4}\int_0^\infty {dx\,x^3\over e^x+1}.
\end{equation}
But
\begin{equation}
\int_0^\infty{x^{s-1}\,dx\over e^x+1}=(1-2^{1-s})\zeta(s)\Gamma(s),
\end{equation}
so here we find
\begin{equation}
f_F=-{7\pi^2\over 1920 a^4},
\end{equation}
which is, indeed, 
 ${7\over4}$ times the scalar force given in Eq.~(\ref{casforce}). 

The effect of fermion fluctuations was first investigated by Ken Johnson
\cite{johnson}, in connection with the bag model \cite{bag}.

\section{Casimir Effect in Dielectrics}
\label{sec:dielectrics}
\setcounter{equation}{0}%

The formalism given in Sec.~\ref{sec:em} can be readily extended to
dielectric bodies \cite{sch1}.  The starting point is the effective
action in the presence of an external polarization source $\bf P$:
\begin{equation}
W=\int(dx)[{\bf P\cdot(-\dot A-\mbox{\boldmath{$\nabla$}}\phi)+\epsilon
E\cdot(-\dot A-\mbox{\boldmath{$\nabla$}}\phi)
-H\cdot(\mbox{\boldmath{$\nabla$}}
\times A)}+{1\over2}(H^2-\epsilon E^2)],
\end{equation}
which, upon variation with respect to $\bf H$, $\bf E$, $\bf A$, and $\phi$,
yields the appropriate Maxwell's equations.  Thus, because $W$ is stationary
with respect to these field variations, its response to a change in dielectric
constant is explicit:
\begin{equation}
\delta_\epsilon W=\int (dx) \delta\epsilon{1\over2}E^2.
\label{dew}
\end{equation}
Comparison of $i\delta_\epsilon W$ with the second iteration of the 
source term in the vacuum persistence amplitude, 
\begin{equation}
e^{iW}=\cdots +{1\over2}\left[i\int (dx) {\bf E\cdot P}\right]^2+\cdots,
\end{equation}
allows us to identify the effective product of polarization sources,
\begin{equation}
i{\bf P}(x){\bf P}(x')\big|_{\rm eff}={\bf 1}\delta\epsilon\,\delta(x-x').
\end{equation}
Thus, the numerical value of the action according to Eq.~(\ref{egreen}),
\begin{equation}
W={1\over2}\int(dx){\bf P}(x)\cdot{\bf E}(x)={1\over2}\int(dx)(dx')
{\bf P}(x)\cdot\mbox{\boldmath{$\Gamma$}}(x,x')\cdot{\bf P}(x'),
\end{equation}
implies the following change in the action when the dielectric constant
is varied slightly,
\begin{equation}
\delta W=-{i\over2}\int (dx)\delta\epsilon(x)\Gamma_{kk}(x,x),
\label{dw}
\end{equation}
which, in view of Eq.~(\ref{dew}),
is equivalent to the vacuum-expectation-value replacement (\ref{ereplace}).

For the geometry of parallel dielectric slabs,  like that considered in
Sec.~\ref{sec:parallel}, but with dielectric materials in the three
regions,
\begin{equation}
\epsilon(z)=\left\{\begin{array}{lc}
\epsilon_1,\quad z<0,\\
\epsilon_3,\quad 0<z<a,\\
\epsilon_2,\quad a<z,
\end{array}\right.
\end{equation}
the components of the Green's dyadics may be expressed in terms of the
TE (transverse electric or H) modes and the TM (transverse magnetic or E)
modes, given by the scalar Green's functions satisfying
\alpheqn
\begin{eqnarray}
\left(-{\partial^2\over\partial z^2}+k^2-\omega^2\epsilon\right)g^E(z,z')
&=&\delta(z-z'),\\
\left(-{\partial\over\partial z}{1\over\epsilon}{\partial\over\partial z'}
+{k^2\over\epsilon}-\omega^2\right)g^H(z,z')
&=&\delta(z-z'),
\end{eqnarray}
\reseteqn
where, quite generally, $\epsilon=\epsilon(z)$, $\epsilon'=\epsilon(z')$.
The nonzero components of the Fourier transform $\bf g$ given by 
Eq.~(\ref{ftgreen}) are easily found to be
(in the coordinate system where $\bf k$ lies along the $+x$ axis)
\alpheqn
\begin{eqnarray}
g_{xx}&=&-{1\over\epsilon}\delta(z-z')+{1\over\epsilon}{\partial\over
\partial z}{1\over\epsilon'}{\partial\over\partial z'}g^H,\label{gxxdi}\\
g_{yy}&=&\omega^2g^E,\\
g_{zz}&=&-{1\over\epsilon}\delta(z-z')+{k^2\over\epsilon\epsilon'}g^H,\\
g_{xz}&=&i{k\over\epsilon\epsilon'}{\partial\over\partial z}g^H,\\
g_{zx}&=&-i{k\over\epsilon\epsilon'}{\partial\over\partial z'}g^H\label{gzxdi}.
\end{eqnarray}
\reseteqn
The trace required in the change of the action (\ref{dw}) is
obtained by taking the limit $z'\to z$, and consequently omitting
delta functions:
\begin{equation}
g_{kk}=\left(\omega^2g^E+{k^2\over\epsilon\epsilon'}g^H
+{1\over\epsilon}{\partial\over\partial z}{1\over\epsilon'}
{\partial\over\partial z'}g^H\right)\bigg|_{z=z'}.
\end{equation}
This appears in the change of the energy when the second interface is
displaced by an amount $\delta a$,
\begin{equation}
\delta\epsilon(z)=-\delta a(\epsilon_2-\epsilon_3)\delta(z-a),
\end{equation}
namely ($A$ is the transverse area)
\begin{equation}
{\delta E\over A}={i\over2}\int{d\omega\over2\pi}{(d{\bf k})\over
(2\pi)^2} dz\,\delta\epsilon(z) g_{kk}(z,z';{\bf k},\omega).
\label{deltaeovera}
\end{equation}
Because $g^E$, $g^H$ and ${1\over\epsilon}{\partial\over\partial z}
{1\over\epsilon'}{\partial\over\partial z'}g^H$ are all continuous, while
$\epsilon\epsilon'$ is not, we interpret the trace of $\bf g$ in 
Eq.~(\ref{deltaeovera}) symmetrically; we let $z$ and $z'$ approach
the interface from opposite sides, so the term ${k^2\over\epsilon\epsilon'}
g^H\to{k^2\over\epsilon_1\epsilon_2}g^H$.  Subsequently, we may evaluate
the Green's function on a single side of the interface.
In terms of the notation
\begin{equation}
\kappa^2=k^2-\omega^2\epsilon,
\end{equation}
which is positive after a complex frequency rotation is performed
(it is automatically positive for finite temperature), the electric
Green's function is in the  region $z,z'>a$,
\begin{equation}
g^E(z,z')={1\over2\kappa_2}\left(e^{-\kappa_2|z-z'|}
+re^{-\kappa_2(z+z'-2a)}\right),
\end{equation}
where the reflection coefficient is
\begin{equation}
r={\kappa_2-\kappa_3\over\kappa_2+\kappa_3}+{4\kappa_2\kappa_3\over
\kappa_3^2-\kappa_2^2}d^{-1},
\end{equation}
with the denominator here being
\begin{equation}
d={\kappa_3+\kappa_1\over\kappa_3-\kappa_1}
{\kappa_3+\kappa_2\over\kappa_3-\kappa_2}e^{2\kappa_3 a}-1.
\end{equation}
The magnetic Green's function $g^H$ has the same form but with the replacement
\begin{equation}
\kappa\to\kappa/\epsilon\equiv\kappa',
\end{equation}
except in the exponentials; the corresponding denominator is denoted by $d'$.
[It is easy to see that $g^E$ reduces to Eq.~(\ref{scgreenout}) when
$r=-1$; the results in Sec.~2.5 follow from Eqs.~(\ref{gxxdi})--(\ref{gzxdi})
in the coordinate system adopted here.]

Evaluating these Green's functions just outside the interface, we find for
the force on the surface per unit area
\begin{eqnarray} 
f={i\over2}\int{d\omega\over2\pi}{(d{\bf k_\perp})\over(2\pi)^2}
\left\{\left[\kappa_3-\kappa_2+2\kappa_3d^{-1}\right]
+\left[\kappa_3-\kappa_2+2\kappa_3d^{\prime-1}
\right]\right\},
\end{eqnarray} where the first bracket comes from the E modes, and the second
from the H modes.  The first term in each bracket, which does not make
reference to the separation $a$ between the surfaces, is seen to be a
change in the volume energy of the system. These terms correspond to the
electromagnetic energy required to replace medium 2 by medium 3 in the
displacement volume.  They constitute the so-called bulk energy
contribution.  The remaining terms are the Casimir force.  We rewrite the
latter by making a complex rotation in the frequency,
\begin{equation}
\omega\to i\zeta, \quad \zeta \mbox{ real}.
\end{equation}
This gives for the force per unit area at zero temperature
\begin{eqnarray}
f^{T=0}_{\rm Casimir}=-{1\over8\pi^2}\int_0^\infty d\zeta \int_0^\infty
dk^2\,2\kappa_3 \left(d^{-1}+d^{\prime-1}
\right),\quad \kappa^2=k^2+\epsilon\zeta^2.
\label{dielectricforce}
\end{eqnarray}
From this, we can obtain the finite temperature expression immediately
by the substitution
\begin{equation}
\zeta^2\to\zeta_n^2=4\pi^2n^2/\beta^2,
\end{equation}
\begin{equation}
\int_0^\infty {d\zeta\over2\pi}\to{1\over\beta}{\sum_{n=0}^\infty}{}',
\end{equation}
the prime being a reminder to count the $n=0$ term with half weight.
These results agree with those of Lifshitz et al. 
\cite{lifshitz,dzyaloshinskii}.

Note that the same result (\ref{dielectricforce})
may be easily rederived by computing the
normal-normal component of the stress tensor on the surface, $T_{zz}$, 
provided two
constant stresses are removed, terms which would be present if either
medium filled all space.  The difference between these two constant 
stresses,
\begin{equation}
T_{zz}^{\rm vol}=-i\int{(d{\bf k_\perp})\over(2\pi)^2}{d\omega\over2\pi}
[\kappa_2-\kappa_3],
\end{equation}
precisely corresponds to the deleted volume energy in the previous
calculation.

Various applications can be made from this general formula
(\ref{dielectricforce}).  In particular, if we set the intermediate
material to be vacuum, $\epsilon_3=1$, and set $\kappa_1=\kappa_2=\infty$,
$\kappa_1'=\kappa_2'=0$, we recover the Casimir force between parallel,
perfectly conducting plates, Eq.~(\ref{casimirclassic}), including the
correct temperature dependence, given by Eq.~(\ref{generaltemp}) with 
$d=2$ and an additional factor of two representing the two polarizations.
Another interesting result, important for the recent experiments \cite{roy},
is the correction for an imperfect conductor, where for frequencies above
the infrared, an adequate representation for the dielectric constant is
\cite{ce}
\begin{equation}
\epsilon(\omega)=1-{\omega_p^2\over\omega^2},
\end{equation}
where the plasma frequency is\footnote{Although we have used rationalized
Heaviside-Lorentz units in our electromagnetic action formalism, that is
without effect, in that the one-loop Casimir effect is independent of
electromagnetic units.  For considerations where the electric charge
and polarizability appear, it seems more convenient to use unrationalized
Gaussian units.}
\begin{equation}
\omega_p^2={4\pi e^2 N\over m},
\end{equation}
where $e$ and $m$ are the charge and mass of the electron, and $N$ is the
number density of free electrons in the conductor.  A simple calculation
shows, at zero temperature \cite{hargreaves,sch1},
\begin{equation}
f\approx-{\pi^2\over240 a^4}\left[1-{8\over3\sqrt{\pi}}{1\over ea}\left(
m\over N\right)^{1/2}\right].
\end{equation}

Now suppose the central slab consists of a tenuous medium and the surrounding
medium is vacuum, so that the dielectric constant in the slab
differs only slightly from unity,
\begin{equation}
\epsilon-1\ll1.
\end{equation}
Then, with a simple change of variable,
\begin{equation}
\kappa=\zeta p,
\end{equation}
we can recast the Lifshitz formula (\ref{dielectricforce}) into the
form
\begin{equation}
f\approx-{1\over32\pi^2}\int_0^\infty d\zeta\,\zeta^3\int_1^\infty
{dp\over p^2}[\epsilon(\zeta)-1]^2[(2p^2-1)^2+1]e^{-2\zeta pa}.
\label{weaklif}
\end{equation}
If the separation of the surfaces is large compared to the
characteristic
wavelength characterizing $\epsilon$, $a\zeta_c\gg1$, we can
disregard
the frequency dependence of the dielectric constant,
and we find
\begin{equation}
f\approx-{23(\epsilon-1)^2\over640\pi^2a^4}.
\label{longdist}
\end{equation}
For short distances, $a\zeta_c\ll1$, the approximation is
\begin{equation}
f\approx-{1\over32\pi^2}{1\over a^3}\int_0^\infty
d\zeta(\epsilon(\zeta)-1)^2.
\label{shortdist}
\end{equation}
These formulas are identical with the well-known forces
 found for the complementary geometry in \cite{sch1}.

Now we wish to obtain these results from the sum of van der Waals
forces,
derivable from a potential of the form
\begin{equation}
V=-{B\over r^\gamma}.
\end{equation}
We do this by computing the energy ($N= $ density of molecules)
\begin{equation}
E=-{1\over2}B N^2\int_0^a dz\int_0^a dz'\int(d{\bf r_\perp})(d{\bf
r'_\perp})
{1\over[({\bf r_\perp-r'_\perp})^2+(z-z')^2]^{\gamma/2}}.
\end{equation}
If we disregard the infinite self-interaction terms (analogous to
dropping the volume energy terms in the Casimir calculation), we
get
\begin{equation}
f=-{\partial\over\partial a}{E\over A}=-{2\pi B
N^2\over(2-\gamma)(3-\gamma)}
{1\over a^{\gamma-3}}.
\end{equation}
So then, upon comparison with (\ref{longdist}), we set $\gamma=7$
and in terms of the polarizability,
\begin{equation}
\alpha={\epsilon-1\over4\pi N},
\end{equation}
we find
\begin{equation}
B={23\over4\pi}\alpha^2,
\label{bee}
\end{equation}
or, equivalently, we recover the retarded dispersion potential
\cite{CasimirandPolder},
\begin{equation}
V=-{23\over4\pi}{\alpha^2\over r^7},
\label{caspol}
\end{equation}
whereas for short distances we recover from Eq.~(\ref{shortdist})
the London potential \cite{London},
\begin{equation}
V=-{3\over\pi}{1\over r^6}\int_0^\infty d\zeta\,\alpha(\zeta)^2,
\end{equation}
which are the quantititive forms of Eqs.~(\ref{retdis}) and (\ref{london}),
respectively.

\section{Casimir Effect on a Sphere}
\label{sec:sphere}
\setcounter{equation}{0}%

\subsection{Perfectly Conducting Spherical Shell}
The zero-point fluctuations due to parallel plates, either conducting or
insulating, give rise to an attractive force, which seems intuitively
understandable in view of the close connection with the attractive
van der Waals interactions.  However, one's intuition fails when more
complicated geometries are considered.

In 1956 Casimir proposed that the zero-point force could be the Poincar\'e
stress stabilizing a semiclassical model of an electron \cite{casimir2}.
For definiteness, take a naive model of an electron as a perfectly
conducting shell of radius $a$ carrying a total charge $e$.  The Coulomb
repulsion must be balanced by some attractive force; Casimir proposed that
that could be the vacuum fluctuation energy, so that the effective energy
of the configuration would be
\begin{equation}
E={e^2\over2 a}-{Z\over a}\hbar c,
\end{equation}
where the Casimir energy is characterized by a pure number $Z$.  The would
open the way for a semiclassical calculation of the fine-structure constant,
for stability results if $E=0$ or
\begin{equation}
\alpha={e^2\over\hbar c}=2Z.
\end{equation}
Unfortunately as Tim Boyer was to discover a decade later
\cite{boyer}, the Casimir force
in this case is {\em repulsive}, $Z=-0.04618$.  The sign results from
delicate cancellations between interior and exterior modes, and between
TE and TM modes, so it appears impossible to predict the sign {\it a priori}.

Boyer's calculation was rather complicated, involving finding the
zeroes of Bessel functions.  In 1978 we found a simpler approach, based
on Green's functions \cite{sch2}, which I will describe here.
In particular the Green's dyadic formalism of Sec.~\ref{sec:em} may be used, 
except now the modes
must be described by vector spherical harmonics, defined by \cite{ce,stratton}
\begin{equation}
{\bf X}_{lm}=[l(l+1)]^{-1/2}{\bf L}Y_{lm}(\theta,\phi),
\end{equation}
where $\bf L$ is the orbital angular momentum operator,
\begin{equation}
{\bf L}={1\over i}{\bf r}\times\mbox{\boldmath{$\nabla$}}.
\end{equation}
The vector spherical harmonics satify the orthonormality condition
\begin{equation}
\int d\Omega\, {\bf X}^*_{l'm'}\cdot {\bf X}^{\vphantom{*}}_{lm}
=\delta_{ll'}\delta_{mm'},
\end{equation}
as well as the sum rule
\begin{equation}
\sum_{m=-l}^l|{\bf X}_{lm}(\theta,\phi)|^2={2l+1\over 4\pi}.
\end{equation}
The divergenceless dyadics $\bGamma' $ and $\bPhi$ may be expanded in
terms of vector spherical harmonics as
\alpheqn
\begin{eqnarray}
\bGamma'&=&\sum_{lm}\left(f_l{\bf X}_{lm}+{i\over\omega}\bnabla\times g_l
{\bf X}_{lm}\right),\\
\bPhi&=&\sum_{lm}\left(\tilde g_l{\bf X}_{lm}-{i\over\omega}\bnabla\times 
\tilde f_l{\bf X}_{lm}\right),
\end{eqnarray}
\reseteqn
where the second suppressed tensor index is carried by the coefficient functions
$f_l$, $g_l$, $\tilde f_l$, $\tilde g_l$.

Inserting this expansion into the first-order equations (\ref{max1})
and (\ref{max2}), and using the properties of the vector spherical
harmonics, we straightforwardly find \cite{sch2} that the Green's dyadic
may be expressed in terms of two scalar Green's functions, the electric
and the magnetic:
\begin{eqnarray}
\bGamma({\bf r,r'};\omega)&=&\sum_{lm}\{\omega^2
F_l(r,r'){\bf X}^{\vphantom{*}}_{lm}(\Omega){\bf X}_{lm}^*(\Omega')\nonumber\\
&&\mbox{}-\bnabla
\times[G_l(r,r'){\bf X}^{\vphantom{*}}_{lm}(\Omega)
{\bf X}_{lm}^*(\Omega')]\times
\overleftarrow\bnabla\}\nonumber\\
&&\mbox{}+\delta\mbox{-function terms},
\end{eqnarray}
where the expression ``$\delta$-function terms'' refers to terms proportional
to spatial delta functions.  These terms may be omitted, as we are interested
in the {\em limit\/} in which the two spatial points approach coincidence.
These scalar Green's functions satisfy the differential equation
\begin{equation}
\left({1\over r}{d^2\over dr^2}r-{l(l+1)\over r^2}+\omega^2\right)
\left\{\begin{array}{c}
F_l(r,r')\\G_l(r,r')
\end{array}\right\}=-{1\over r^2}\delta(r-r'),
\end{equation}
subject to the boundary conditions that they be finite at the origin (the
center of the sphere), which picks out $j_l$ there, and correspond to
outgoing spherical waves at infinity, which picks out $h_l^{(1)}$.  On the
surface of the sphere, we must have
\begin{equation}
F_l(a,r')=0,\quad {\partial\over\partial r}rG_l(r,r')\bigg|_{r=a}=0.
\end{equation}
The result is that
\begin{equation}
\left\{\begin{array}{c}
G_l\\F_l\end{array}\right\}=G_l^0+\left\{\begin{array}{c}
\tilde G_l\\\tilde F_l\end{array}\right\},
\end{equation}
where $G_l^0$ is the vacuum Green's function,
\begin{equation}
G_l^0(r,r')=ikj_l(kr_<)h_l^{(1)}(kr_>),
\end{equation}
and in the interior and the exterior of the sphere respectively,
\alpheqn
\begin{eqnarray}
r,r'<a:\quad\left\{\begin{array}{c}
\tilde G_l\\\tilde F_l\end{array}\right\}&=&-A_{G,F}ikj_l(kr)j_l(kr'),\\
r,r'>a:\quad\left\{\begin{array}{c}
\tilde G_l\\
\tilde F_l\end{array}\right\}&=&-B_{G,F}ikh^{(1)}_l(kr)h^{(1)}_l(kr'),
\end{eqnarray}
\reseteqn
where the coefficients are
\alpheqn
\begin{eqnarray}
A_F=B_F^{-1}={h_l^{(1)}(ka)\over j_l(ka)},\\
A_G=B_G^{-1}={[kah_l^{(1)}(ka)]'\over[ kaj_l(ka)]'},
\end{eqnarray}
\reseteqn
From the electromagnetic energy density we may derive the following
formula for the energy of the system
\begin{eqnarray}
E&=&\int(d{\bf r}){1\over2i}\int_{-\infty}^\infty {d\omega\over2\pi}
e^{-i\omega(t-t')}\sum_{lm}\{k^2[\tilde F_l(r,r')+\tilde G_l(r,r')]
{\bf X}^{\vphantom{*}}_{lm}(\Omega)\cdot{\bf X}^*_{lm}(\Omega')\nonumber\\
&&\mbox{}-\bnabla\times{\bf X}^{\vphantom{*}}_{lm}
(\Omega)\cdot[\tilde F_l(r,r')
+\tilde G_l(r,r')]\cdot{\bf X}_{lm}^*(\Omega')\times\overleftarrow\bnabla\}
\bigg|_{{\bf r=r'}}.
\end{eqnarray} Note here that the vacuum term in the Green's
functions has been removed, since that corresponds to the zero-point energy
that would be present in this formalism if no bounding surface were
present.  Here we are putting the two spatial points coincident, while
we leave a temporal separation, $\tau=t-t'$, which is only to be set equal
to zero at the end of the calculation, and therefore serves as a kind of
regulator.  The integration over the solid angle and the sum on  $m$ may 
be easily carried out, and the radial integral over Bessel functions is
simply done using recurrence relations.
After performing a complex frequency rotation,
we obtain the following compact form for the Casimir energy:
\begin{equation}
E=-{1\over2\pi a}\sum_{l=1}^\infty(2l+1){1\over2}\int_{-\infty}^\infty
dy\,e^{i\epsilon y}x{d\over dx}\ln(1-\lambda_l^2),
\end{equation}
where \begin{equation}
\lambda_l=[s_l(x)e_l(x)]'
\end{equation}
is written in terms of Ricatti-Bessel functions of imaginary argument,
\begin{eqnarray}
s_l(x)&=&\sqrt{\pi x\over2}I_{l+1/2}(x),\nonumber\\
e_l(x)&=&\sqrt{2x\over\pi}K_{l+1/2}(x).
\end{eqnarray}
Here, as a result of the Euclidean rotation,
\begin{equation}
x=|y|,\quad y={1\over i}ka \mbox{ is real, as is}\quad
\epsilon={1\over i}{\tau\over a}\to 0.
\end{equation}
The same formula may be derived by computing the stress on the surface
through use of the stress tensor \cite{sch2}.

A very rapidly convergent evaluation of this formula can be achieved
by using the uniform asymptotic expansions for the Bessel functions:
\alpheqn
\begin{eqnarray}
s_l(x)&\sim&{1\over2}{z^{1/2}\over(1+z^2)^{1/2}}e^{\nu \eta}\left[
1+\sum_{k=1}^\infty{u_k(t)\over\nu^k}\right],\\
e_l(x)&\sim&{z^{1/2}\over(1+z^2)^{1/2}}e^{-\nu \eta}\left[
1+\sum_{k=1}^\infty{(-1)^ku_k(t)\over\nu^k}\right],\quad l\to\infty,
\end{eqnarray}
\reseteqn
where \begin{equation}
x=\nu z,\, \nu=l+1/2,\, t=(1+z^2)^{-1/2},\, \eta=(1+z^2)^{1/2}
+\ln{z\over1+(1+z^2)^{1/2}},
\end{equation}
and the $u_k(t)$ are polynomials in $t$ of definite
parity and of order $3k$ \cite{stegun}.
If we now write
\begin{equation}
E=-{1\over2a}\sum_{l=1}^\infty J(l,\epsilon),
\end{equation}
we easily find
\begin{equation}
J(l,0)\sim{3\over32}, \quad l\to\infty.
\end{equation}
In order to obtain a finite sum, therefore, we must keep $\epsilon\ne0$ until
the end of 
the calculation.  By adding and subtracting the leading approximation
to the logarithm, we can write
\begin{equation}
J(l,\epsilon)=R_l+S_l(\epsilon),
\end{equation}
where
\begin{equation} 
R_l=-{1\over2\pi}\int_0^\infty dz\left[(2l+1)^2\ln(1-\lambda_l^2)+{1\over(1+
z^2)^3}\right]=J(l,0)-{3\over32},
\end{equation}
and 
\begin{equation}
S_l(\epsilon)=-{1\over4\pi}\int_{-\infty}^\infty z\,dz\,e^{i\epsilon \nu z}
{d\over dz}{1\over(1+z^2)^3}.
\end{equation}
By use of the Euler-Maclaurin sum formula (\ref{eulermac}), 
we can work out the sum
\begin{equation}
\sum_{l=1}^\infty S_l(\epsilon)=-{3\over32},
\end{equation}
precisely the negative of the value of a single term at $\epsilon=0$!
The sum of the remainder, $\sum_l R_l$, is easily evaluated numerically,
and changes this result by less than 2\%.  Thus the result for the
Casimir energy for a spherical conducting shell is found to be
\begin{equation}
E={0.092353\over2a}.
\label{boyerresult}
\end{equation}
This agrees with the result found in 1968 by Boyer \cite{boyer}, evaluated
more precisely by Davies \cite{davies}, and confirmed by a completely
different method by Balian and Duplantier in 1978 \cite{balian}.
Recently, this result has been reconfirmed, using a zeta function method,
by Leseduarte and Romeo \cite{romeo}.  A reconsideration using direct
mode summation has appeared very recently \cite{hagen}.

\subsection{Fermion Fluctuations}
The corresponding calculation for a massless 
spin-1/2 particle subject to bag model
boundary conditions (\ref{bagbc}) on a spherical surface,
\begin{equation}
(1+i{\bf n}\cdot\mbox{\boldmath{$\gamma$}})G\bigg|_{S}=0,
\label{bmbc}
\end{equation}
 was carried out by Ken Johnson \cite{ken2} and by Milton
\cite{fermion}.  The result is also a repulsive stress, of less than one-half
the magnitude of the electromagnetic result.  (Recall that for parallel
plates, the reduction factor was $7/8$.)

In this case we wish to solve the Green's function equation
\begin{equation}
\left(\gamma{1\over i}\partial\right)G(x,x')=\delta(x-x')
\end{equation}
subject to the boundary condition (\ref{bmbc}).  In the same representation
for the gamma matrices used before, this may be easily achieved in terms
of the total angular momentum eigenstates (${\bf J=L}+(1/2)
\mbox{\boldmath{$\sigma$}}$):
\begin{eqnarray}
Z_{JM}^{l=J\pm1/2}(\Omega)&=&\left(l+1/2\mp M\over2l+1\right)^{1/2}Y_{lM-1/2}
(\Omega)|+\rangle\nonumber\\
&&\quad\mbox{}\mp\left(l+1/2\pm M\over2l+1\right)^{1/2}Y_{lM+1/2}
(\Omega)|-\rangle.
\end{eqnarray}
These satisfy
\begin{equation}
\mbox{\boldmath{$\sigma$}}\cdot{\bf \hat r}Z_{JM}^{l=J\pm1/2}=
Z_{JM}^{l=J\mp1/2}.
\end{equation}  Once the Green's function is found, it can be used in
the usual way to compute the vacuum expectation value of the stress tensor,
which in the Dirac case is given by
\begin{equation}
T^{\mu\nu}={1\over2}\psi\gamma^0{1\over2}\left(\gamma^\mu{1\over i}\partial^\nu
+\gamma^\nu{1\over i}\partial^\mu\right)\psi+g^{\mu\nu}{\cal L},
\end{equation}
${\cal L}$ being the fermionic Lagrange function,
which leads directly to
\begin{equation}
T_{rr}={1\over2}{\partial\over \partial r}\mbox{tr}
\,\mbox{\boldmath{$\gamma$}}
\cdot{\bf \hat r}G(x,x')\bigg|_{x'\to x}.
\end{equation}
The discontinuity of the stress tensor across the surface of the sphere
gives the energy according to
\begin{equation}
4\pi a^2[T_{rr}(a-)-T_{rr}(a+)]=-{\partial\over\partial a}E(a).
\end{equation}
The result of a quite straightforward calculation (the details are given in
Ref.~\cite{milton80})
gives the result for the
sum of exterior and exterior modes, again, in terms of modified spherical
Bessel functions:
\begin{equation}
E=-{2\over\pi a}\sum_{l=0}^\infty(l+1)\int_0^\infty dx\,x\cos x\epsilon\,
{d\over dx}\ln[(e_l^2+e_{l+1}^2)(s_l^2+s_{l+1}^2)].
\end{equation}
This may again be numerically evaluated through use of the uniform
asymptotic approximants, with the result
\begin{equation}
E={0.0204\over a}.
\end{equation}
Somewhat less precison was obtained because, in this case, the leading
uniform asymptotic approximation vanished.

\section{Infinite Circular Cylinder}
\setcounter{equation}{0}%

Since parallel plates yield an attractive Casimir force, and a sphere has
a repulsive stress, one might guess that for a cylinder a zero stress results.
The situation is not so simple.  The first calculation was carried out in
1981 by DeRaad and Milton \cite{cylinder}.
The electrodynamic result turns out to be attractive but with rather 
small magnitude.

Consider a right circular perfectly conducting
cylinder of infinite length and radius $a$. We compute the Casimir energy
using the above Green's dyadic formalism, adapted to this cylindrical
basis.  The necessary information about vector spherical harmonics
in this case is given in Stratton \cite{stratton}.  The results for the
Green's dyadics are
\alpheqn
\begin{eqnarray}
\bGamma({\bf r,r'})&=&\sum_{m=-\infty}^\infty\int{dk\over2\pi}\bigg[
-{1\over\omega^2}{\bf MM^{\prime*}}(d_m-k^2)F_m(r,r')\nonumber\\
&&\mbox{}+{1\over\omega^2}
{\bf NN^{\prime*}}G_m(r,r')\bigg]\chi^{\vphantom{*}}_{mk}(\theta,z)
\chi^*_{mk}(\theta',z'),\\
\bPhi({\bf r,r'})&=&\sum_{m=-\infty}^\infty\int{dk\over2\pi}\bigg[
-i{\bf MN^{\prime*}}G_m(r,r')\nonumber\\
&&\mbox{}-{i\over\omega}
{\bf NM^{\prime*}}F_m(r,r')\bigg]\chi^{\vphantom{*}}_{mk}(\theta,z)
\chi^*_{mk}(\theta',z'),
\end{eqnarray}
\reseteqn
where the nonradial eigenfunctions are
\begin{equation}
\chi_{mk}(\theta,z)={1\over\sqrt{2\pi}}e^{im\theta}e^{ikz},
\end{equation}
the vector differential operators $\bf M$ and $\bf N$ are
\begin{equation}
{\bf M}={\bf \hat r}{im\over r}-\mbox{\boldmath{$\hat\theta$}}{\partial\over
\partial r},\quad {\bf N}={\bf \hat r}ik{\partial\over\partial r}
-\mbox{\boldmath{$\hat\theta$}}{mk\over r}-{\bf \hat z}d_m,
\end{equation}
and the scalar differential operator $d_m$ is
\begin{equation}
d_m={1\over r}{\partial\over\partial r}r{\partial\over\partial r}-
{m^2\over r^2}.
\end{equation}
The scalar Green's functions in the interior and exterior regions are
\alpheqn
\begin{eqnarray}
r,r'<a:&&\nonumber\\
{1\over\omega^2}F_m(r,r')&=&{i\pi\over2\lambda^2}\left[J_m(\lambda r_<)
H_m(\lambda r_>)-{H_m'(\lambda a)\over J'_m(\lambda a)}J_m(\lambda r)
J_m(\lambda r')\right]\nonumber\\
&&\mbox{}+{1\over\lambda^2}{\cal G}_m^F(r,r'),\\
{1\over\omega}G_m(r,r')&=&{i\pi\over2\lambda^2}\left[J_m(\lambda r_<)
H_m(\lambda r_>)-{H_m(\lambda a)\over J_m(\lambda a)}J_m(\lambda r)
J_m(\lambda r')\right]\nonumber\\
&&\mbox{}+{1\over\lambda^2}{\cal G}_m^G(r,r'),\\
{\cal G}_m^{G,F}(r,r')&=&-{1\over2|m|}\left(r_<\over r_>\right)^{|m|}
\pm{1\over2|m|}{r^{|m|}r^{\prime|m|}\over a^{2|m|}},\quad m\ne0,
\end{eqnarray}
\reseteqn
\alpheqn
\begin{eqnarray}
r,r'>a:&&\nonumber\\
{1\over\omega^2}F_m(r,r')&=&{i\pi\over2\lambda^2}\left[J_m(\lambda r_<)
H_m(\lambda r_>)-{J_m'(\lambda a)\over H'_m(\lambda a)}H_m(\lambda r)
H_m(\lambda r')\right]\nonumber\\
&&\mbox{}+{1\over\lambda^2}{\cal G}_m^F(r,r'),\\
{1\over\omega}G_m(r,r')&=&{i\pi\over2\lambda^2}\left[J_m(\lambda r_<)
H_m(\lambda r_>)-{J_m(\lambda a)\over H_m(\lambda a)}H_m(\lambda r)
H_m(\lambda r')\right]\nonumber\\
&&\mbox{}+{1\over\lambda^2}{\cal G}_m^G(r,r'),\\
{\cal G}_m^{G,F}(r,r')&=&-{1\over2|m|}\left(r_<\over r_>\right)^{|m|}
\pm{1\over2|m|}{a^{2|m|}\over r^{|m|}r^{\prime|m|}},\quad m\ne0,
\end{eqnarray}
\reseteqn
where $H_m=H_m^{(1)}$ is the Hankel function of the first kind, and $\lambda^2
=\omega^2-k^2$.  Although ${\cal G}^{G,F}_0$ are not determined, they do not
contribute to physical quantities.

The result for the force per unit area is ($z=\lambda a$)
\begin{eqnarray}
&&f=T_{rr}(a_-)-T_{rr}(a_+)\nonumber\\
&&\!\!\!\!\!\!\!=-{1\over2ia^2}\int_{-\infty}^\infty{d\omega\over2\pi}
\psi(\omega){1\over2\pi}
\sum_{m=-\infty}^\infty\int_{-\infty}^\infty{dk\over2\pi}
\bigg[z{\partial\over\partial z}
\ln\left(1-{\pi\over4}z^2[(H_m(z)J_m(z))']^2\right)\bigg].
\end{eqnarray}Here, because convergence is more subtle than in the previous
cases, we have inserted a frequency cutoff function $\psi(\omega)$ that
has the properties
\begin{eqnarray}
\psi(0)&=&1,\nonumber\\
\psi(\omega)&\sim&{1\over|\omega|^4},\quad |\omega|\to\infty,\nonumber\\
\psi(\omega) &&\mbox{ real for } \omega \mbox{ real or imaginary}.
\end{eqnarray}
A way of satisfying these conditions is to take
\begin{equation}
\psi(\omega)=\sum_i\left({a_i\over a^2\omega^2-\mu_i^2}+{a_i^*\over a^2\omega^2
-\mu_i^*}\right),
\end{equation}
with the conditions on the residues and poles:
\alpheqn
\begin{eqnarray}
\sum_i\mbox{Re}\,a_i&=&0,\\
2\sum_i\mbox{Re}\,{a_i\over\mu_i^2}&=&-1.
\end{eqnarray}
\reseteqn
Of course, the poles are to recede to infinity.  When we rotate the contour
to imaginary frequencies, the contribution of these poles is necessary to
achieve a real result.

Once again we use the uniform asymptotic expansion to extract the leading
behavior.  We evaluate the $m$ sum for the leading term by using the
identity (\ref{cotan}), or
\begin{eqnarray}
2\sum_{m=1}^\infty{1\over m^2\rho-z^2}
&=&-{\pi\over\sqrt{-z^2\rho}}\left(1-\coth\pi
\sqrt{-z^2\over\rho}\right)\nonumber
\\&&\mbox{}+{1\over z^2}+{\pi\over\sqrt{-z^2\rho}}.
\end{eqnarray}
The result for the energy per unit length is
\begin{equation}
{\cal E}=\pi a^2 f=-{1\over8\pi a^2}(S+R+R_0),
\end{equation}
where
\alpheqn
\begin{eqnarray}
S&=&{1\over2}\int_{\epsilon\to0}^\infty dr\,\pi(\coth\pi r-1)\nonumber\\
&&\mbox{}+{1\over2}\left(5{d\over d\rho}+{7\over2}{d^2
\over d\rho^2}+{1\over2}{d^3\over d\rho^3}\right)
\int_0^\infty dr\left[{\pi\over\sqrt{\rho}}\left(\coth{\pi r\over\sqrt{\rho}}
-1\right)-{1\over r}\right]\bigg|_{\rho=1}\nonumber\\
&=&-{1\over2}\ln2\pi\epsilon+{5\over8},\\
R&=&-4\sum_{m=1}^\infty\int_0^\infty r\,dr\left\{\ln\left[1-(r(I_m(r)K_m(r))'
)^2\right]+{r^2\over4(m^2+r^2)^3}\right\}\nonumber\\
&=&-0.0437,\\
R_0&=&-2\int_{\epsilon\to0}^\infty r\,dr\left\{\ln\left[1-(r(I_0(r)K_0(r))')^2
\right]+{1\over4r^2}\right\}-{1\over 4}\nonumber\\
&=&{1\over2}\ln\epsilon+0.6785.
\end{eqnarray}
\reseteqn
Adding these numbers, we see that the infrared singularity $\epsilon$ cancels,
and we obtain an attractive result,
\begin{equation}
{\cal E}=-{0.01356\over a^2}.
\end{equation}

Very recently, this result has been confirmed by two independent
calculations.  First, Gosdzinsky and Romeo obtained the same answer, to
eight significant figures, using
a zeta-function technique \cite{gos}. Shortly thereafter, 
an earlier calculation by
Nesterenko was corrected to yield the same answer \cite{nest}.
This later method is based not on the Green's function, but rather
an analytic evaluation of the sum of zero-point energies through the
formal formula
\begin{equation}
E={1\over2}\int_{-\infty}^\infty {dk\over2\pi}\sum_{m=-\infty}^\infty
{1\over2\pi i}{1\over2}\oint_C\omega\, d_\omega\ln f_n(k,\omega,a),
\end{equation}
where $f_n$ is a function, the zeroes of which are the mode frequencies
of the system, and where $C$ is a contour initially chosen to encircle
those zeroes.  The divergences there are regulated by a zeta function
technique; the results, both analytically and numerical, are identical
to those reported here and derived in Ref.~\cite{cylinder}.

\section{Casimir Effect on a $D$-dimensional Sphere}
\setcounter{equation}{0}%

Because of the rather mysterious dependence of the sign and magnitude
of the Casimir stress on the topology and dimensionality of the
bounding geometry, we have carried out calculation of TE and TM modes bounded
by a spherical shell in $D$ spatial dimensions.  We first consider massless
scalar modes satisfying Dirichlet boundary conditions on the surface, which are
equivalent to electromagnetic TE modes.  Again we calculate the 
vacuum expectation value of the stress on the surface from the Green's 
function.

The Green's function $G({\bf x},t;{\bf x}',t')$ satisfies the inhomogeneous 
Klein-Gordon equation (\ref{kg}), or
\begin{equation}
\left ({{\partial^2}\over{\partial t^2}}-\nabla^2\right ) G({\bf
x},t;{\bf x}',t') =\delta^{(D)} ({\bf x}-{\bf x}')\delta (t-t'),
\label{4}
\end{equation}
where $\nabla^2$ is the Laplacian in $D$ dimensions.
We will solve the above Green's function equation by dividing space
into two regions, {\sl region I}, the interior of a sphere of radius $a$ and
{\sl region II}, the exterior of the sphere. On the sphere we will impose
Dirichlet boundary conditions
\begin{equation}
G({\bf x},t;{\bf x}',t')\bigm | _{|{\bf x}|=a}=0.
\label{5}
\end{equation}
In addition, in  region I we will require that $G$ be finite
at the origin ${\bf x}=0$ and in region II we will require that $G$
satisfy outgoing-wave boundary conditions at $|{\bf x}|=\infty$.

The radial Casimir force per unit area $f$ on the sphere is
obtained from the radial-radial component of the vacuum expectation 
value of the stress-energy tensor \cite{sch2}:
\begin{equation}
f =\langle 0|T^{rr}_{\rm in}-T^{rr}_{\rm out}|0\rangle\bigm |_{r=a}.
\label{7}
\end{equation}
To calculate $f$ we exploit the connection between the vacuum expectation
value of the stress-energy tensor $T^{\mu\nu} ({\bf x},t)$ and the Green's
function at equal times $G({\bf x},t;{\bf x}',t)$:
\begin{equation}
f ={1\over 2i}\left [{\partial\over\partial r}{\partial\over\partial r'}G({\bf
x},t;{\bf x}',t)_{\rm in}-{\partial\over\partial r}{\partial\over\partial r'}
G({\bf x},t;{\bf x}',t)_{\rm out}\right]
\Bigg |_{{\bf x}={\bf x}',~|{\bf x}|=a}.
\label{8}
\end{equation}

To evaluate the expression in (\ref{8}) it is necessary to solve the Green's
function equation (\ref{4}). We begin by taking the time Fourier transform of
$G$:
\begin{equation}
{\cal G}_\omega ({\bf x};{\bf x}')=\int_{-\infty}^{\infty}dt\,e^{i\omega
(t-t')} G({\bf x},t;{\bf x}',t').
\label{9}
\end{equation}
The differential equation satisfied by ${\cal G}_\omega$ is
\begin{equation}
-\left ( \omega^2+\nabla^2 \right ) {\cal G}_\omega ({\bf x};{\bf x}')
= \delta^{(D)} ({\bf x}-{\bf x}').
\label{10}
\end{equation}

To solve this equation we introduce polar coordinates and seek a solution that
has cylindrical symmetry; i.e., we seek a solution that is a function only of
the two variables $r=|{\bf x}|$ and $\theta$, the angle between ${\bf x}$ and
${\bf x}'$ so that ${\bf x}\cdot{\bf x'}= r r'\cos\theta$. In terms of these
polar variables (\ref{10}) becomes
\begin{eqnarray}
&&-\left (\omega^2+{\partial^2\over\partial r^2}
+{D-1\over r}{\partial\over\partial
r}+{\sin^{2-D}\theta\over r^2}{\partial\over\partial\theta}\sin^{D-2}\theta
{\partial\over\partial\theta}\right )
{\cal G}_\omega (r,r',\theta)\nonumber\\
&&\quad=-{\Gamma\left ( {D-1\over 2}\right )\over 2\pi^{(D-1)/2}r^{D-1}
\sin^{D-2} \theta }\delta(r-r')\delta(\theta).
\label{11}
\end{eqnarray}
Note that the $D$-dimensional delta function 
on the right side of (\ref{10}) has
been replaced by a cylindrically-symmetric delta function having the property
that its volume integral in $D$ dimensional space is unity. The $D$-dimensional
volume integral of a cylindrically-symmetric function $f(r,\theta)$ is
\begin{equation}
{2\pi^{(D-1)/2}\over \Gamma\left ( {D-1\over 2}\right )}
\int_0^{\infty}dr\, r^{D-1}\int_0^\pi d\theta \,\sin^{D-2} \theta f(r,\theta).
\label{12}
\end{equation}

We solve (\ref{11}) using the method of separation of variables. Let
\begin{equation}
{\cal G}_\omega (r,r', \theta)= A(r) B(z),
\label{13}
\end{equation}
where $z=\cos\theta$. The equation satisfied by $B(z)$ is then
\begin{equation}
\left [ (1-z^2){d^2\over dz^2}-z(D-1){d\over dz}+n(n+D-2)\right ]B(z)=0,
\label{14}
\end{equation}
where we have anticipated a convenient form for the separation constant. The
equation satisfied by $A(r)$ is
\begin{equation}
\left [ {d^2\over dr^2}+{D-1 \over r}{d\over dr}-{n(n+D-2)\over r^2}+
\omega^2 \right ] A(r)=0 \quad (r\neq r').
\label{15}
\end{equation}
The solution to Eq.~(\ref{14}) that is regular at $|z|=1$ is the ultraspherical
(Gegenbauer) polynomial \cite{stegun}
\begin{equation}
B(z)= C_n ^{(-1+D/2)} (z) \quad (n=0,\,1,\,2,\,3,\,\ldots).
\label{16}
\end{equation}
The solution in region I to Eq.~(\ref{15}) that is regular 
at $r=0$ involves the Bessel function \cite{stegun}
\begin{equation}
A(r)= r^{1-D/2} J_{n-1+{D\over 2}}(|\omega| r).
\label{17}
\end{equation}
In Eq.~(\ref{17}) we assume that $D\geq 2$ in order to eliminate the linearly
independent solution $A(r)=r^{1-D/2}Y_{n-1+{D\over 2}}(|\omega| r)$, which is
singular at $r=0$ for all $n$. The solution in region II to Eq.~(\ref{15})
that corresponds to an outgoing wave at $r=\infty$ involves a Hankel function of
the first kind \cite{stegun}
\begin{equation}
A(r)=r^{1-D/2} H_{n-1+{D\over 2}}^{(1)} (|\omega| r).
\label{18}
\end{equation}

Using a few properties of the ultraspherical polynomials, namely, 
orthonormality \cite{stegun}
\begin{equation}
\int_{-1}^1 dz\, (1-z^2)^{\alpha -1/2} C_n^{(\alpha)}(z)C_m^{(\alpha)}(z)
={ 2^{1-2\alpha} \pi \Gamma(n+2\alpha) \over
n!\, (n+\alpha) \Gamma^2(\alpha)}\delta_{nm} \quad (\alpha\neq 0) ,
\label{22}
\end{equation}
and the value of the ultraspherical polynomials at $z=1$,
\begin{equation}
C_n^{(\alpha)}(1)={\Gamma(n+2\alpha)\over n!\,\Gamma(2\alpha)}
\quad (\alpha\neq 0),
\label{23}
\end{equation}
as well as the duplication formula (\ref{dupform}),
%$\Gamma(2x)=2^{2x-1}\Gamma(x)\Gamma(x+1/2)/
%\sqrt{\pi}$.
we solve for Green's function in the two regions.
Adding the interior and the exterior contributions, and performing the
usual imaginary frequency rotation, we obtain the final expression for
the stress \cite{bender}:
\begin{eqnarray}
f&=&-\sum_{n=0}^{\infty}{(n-1+{D\over 2})\Gamma (n+D-2)\over
2^{D-1}\pi^{D+1\over 2}a^{D+1} n!\,\Gamma\left ({D-1\over 2}\right
)}\nonumber\\
&&\quad\quad\times\int_0^{\infty}dx\left [ x
{d\over dx}\ln\left ( I_{n-1+{D\over 2}}(x)K_{n-1+{D\over2}}(x)x^{2-D}\right )
\right ].
\label{e4}
\end{eqnarray}
It is easy to check that this reduces to the known case at $D=1$, for there
the series truncates---only $n=0$ and 1 contribute, and we easily
find
\begin{equation}
f={F\over2}=-{\pi\over96a^2},
\end{equation}
which agrees with Eq.~(\ref{scalarforce}) for $d=0$ and $a\to2a$.

In general, however, although each $x$ integral can be made finite, the
sum over $n$ still diverges.  We can make the integrals finite by replacing
the $x^{2-D}$ factor in the logarithm by simply $x$, which we are permitted
to do if $D<1$ because
\begin{equation}
\sum_{n=0}^{\infty}{\Gamma(n+\alpha)\over n!}\equiv 0 \quad 
(\alpha<0, \alpha\ne-N).
\label{f1}
\end{equation}
Then the total stress on the sphere is
\begin{equation}
F=-{1\over\pi a^2}\sum_{n=0}^\infty {\Gamma(n+D-2)\over n!\Gamma(D-1)}\nu
Q_\nu\quad \left(\nu=n+{D\over2}-1\right),
\label{totdstress}
\end{equation}
where the integral is
\begin{equation}
Q_\nu=-\int_0^\infty dx\,\ln\left(2xI_\nu(x)K_\nu(x)\right).
\end{equation}
We proceed as follows:
\begin{itemize}
\item Analytically continue to $D<0$, where the sum converges, although the
integrals become complex.
\item Add and subtract the leading asymptotic behavior of the integrals.
\item Continue back to $D>0$, where everything is now finite.
\end{itemize}

We used two methods to carry out the numerical evaluations, which gave
the same results.  
In the first method we use
the uniform asymptotic expansions to yield
\begin{equation}
Q_\nu\sim{\nu\pi\over2}+{\pi\over128\nu}-{35\pi\over32768\nu^3}+{565\pi\over
1048576\nu^5}+\dots.
\end{equation}
Then using the identities
\alpheqn
\begin{eqnarray}
\sum_{n=0}^\infty {\Gamma(n+D-2)\over n!}&=&0 \quad \mbox{for} \,D<2,\\
\sum_{n=0}^\infty {\Gamma(n+D-2)\over n!}\nu^2&=&0 \quad \mbox{for} \,D<0,
\end{eqnarray}
\reseteqn
we obtain the following expression convergent for $D<4$ (further subtractions
can be made for higher dimensions):
\begin{eqnarray}
F&\approx&-{1\over a^2\pi}{1\over\Gamma(D-1)}\biggm\{\sum_{n=0}^{[N-D/2+1]}
{\Gamma(n+D-2)\over n!}\nu\tilde Q_\nu\nonumber\\
&&\mbox{}+\sum_{n=[N-D/2+2]}^\infty{\Gamma(n+D-2)\over n!}\left[
-{35\pi\over32768\nu^2}+{565\pi\over1048576\nu^4}\right]\biggm\},
\end{eqnarray}
where $\tilde Q_\nu=Q_\nu-\nu\pi/2-\pi/128\nu$, and the square
brackets denote the largest integer less or equal to its argument.  
The infinite sums are easily evaluated in terms of gamma functions.

In the second method we carry out an asymptotic expansion of the summand
in (\ref{totdstress}) in $n$, $n\to\infty$
\begin{eqnarray}
{\Gamma(n+D-2)\over n!}\nu Q_\nu&\sim&
{1\over2}\bigg [ n^{D-1} +{(D-1)(D-2)\over2}n^{D-2}\nonumber\\
 \mbox{}&+&{24 D^4 - 176 D^3
+504 D^2 -688 D +387\over 192} n^{D-3} + \ldots \bigg ].
\end{eqnarray}
The sums on $n$ in the terms in the asymptotic expansion are carried out
according to
\begin{eqnarray}
\sum_{n=1}^\infty n^{D-k}=\zeta(k-D),\quad D-k<-1.
\end{eqnarray}
In this way we obtain the following formula suitable for numerical
evaluation:
\begin{eqnarray}
F&=&-{1\over a^2}\biggm\{{1\over2\pi}Q_{D/2-1}\nonumber\\
&&\mbox{}+{1\over\Gamma(D-1)}\sum_{n=1}^\infty\left[{\Gamma(n+D-2)\over n!}
{\nu\over\pi}Q_\nu-{1\over2}\sum_{k=1}^l n^{D-k}c_k(D)\right]\nonumber\\
&&\mbox{}+{1\over2\Gamma(D-1)}\sum_{k=1}^l\zeta(k-D)c_k(D)\biggm\},
\end{eqnarray}where the $c_k(D)$ are polynomials in $D$.

\begin{figure}
\centerline{
\psfig{figure=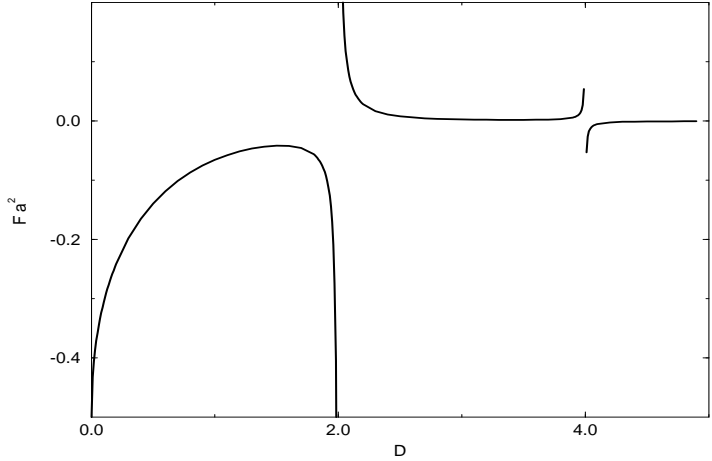,height=4.5in,width=3in,angle=270}}
\caption{Scalar Casimir stress $F$ for $0<D<5$ on a spherical shell.}
\label{figdc}
\end{figure}

Both methods give the same results \cite{bender}, 
which are shown in Fig.~\ref{figdc}.
Note the following salient features:
\begin{itemize}
\item Poles occur at $D=2n$, $n=1,2,3,\dots$.
\item As we will see in the next plot for negative $D$, branch points occur
at $D=-2n$, $n=0,1,2,3,\dots$, and the stress is complex for $D<0$.
\item The stress vanishes at negative even integers, $F(-2n)=0$, 
$n=1,2,3,\dots$, but is nonzero at $D=0$: $F(0)=-1/2a^2$.
\item The case of greatest physical interest, $D=3$, has a finite stress,
but one which is much smaller than the corresponding electrodynamic
one: $F(3)=+0.0028168/a^2$. (This result was confirmed in Ref.~\cite{romeo}.)
\end{itemize}

\begin{figure}
\centerline{
\psfig{figure=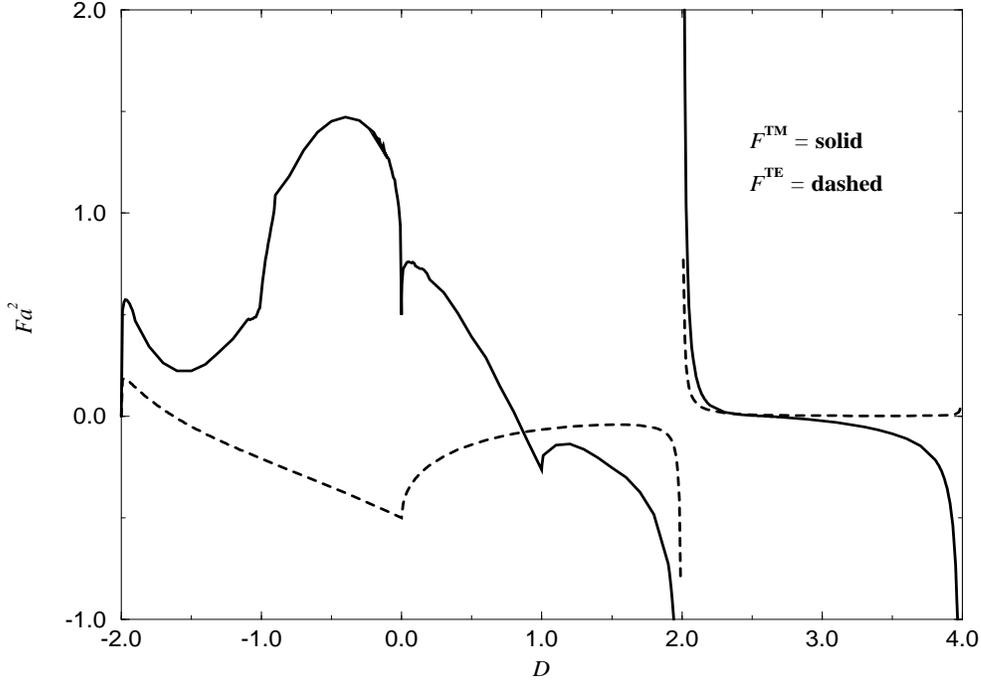,height=6in,width=4in,angle=270}}
\caption{A plot of the TM and TE Casimir stress for $-2<D<4$ on a spherical
shell. For $D<2$ ($D<0$) the stress $F^{\rm TM}$ ($F^{\rm TE}$) is complex
and we have plotted the real part.}  
\label{figvdc}
\end{figure}

The same kind of calculation can be carried out for the TM modes \cite{vdc}.
The TM modes are modes which satisfy mixed boundary conditions on the surface
\cite{boyer,slater},
\begin{equation}
{\partial\over\partial r}r^{D-2}G({\bf x},t;{\bf x}',t')
\bigg|_{|{\bf x}|=r=a}=0,
\label{tmbc}
\end{equation} 
The results are qualitatively similar, and are shown in Fig.~\ref{figvdc}.
In particular, removing the $n=0$
contribution from the sum of the TE and TM contributions, we recover the
Boyer result (\ref{boyerresult}).

\section{Dielectric Ball: A Model for Sonoluminescence?}
\setcounter{equation}{0}%

In all the above illustrations, finiteness of the Casimir energy has
been the result of delicate cancellation between interior and exterior
modes.  This cancellation occurs because the speed of light is identical
in the two regions.  This indicates that if one considered, for example,
a dielectric ball, the procedure given here would not yield a finite
result, and indeed, such is the case.  The same thing occurs for cavity
modes, which are relevant for gluon (or quark) fluctuations in the bag model of
hadrons---for a discussion of these issues see Ref.~\cite{towardfinite}.

A new application of such cavity modes has recently appeared, in
connection with the suggestion by Schwinger \cite{js} that the
dynamical Casimir effect could be relevant for the roughly 1 million
photons emitted per cycle in sonoluminescence \cite{sonorev}.  There
has been considerable controversy concerning this suggestion
\cite{eberlein,carlson,visser}.  The difficulty is that it has been
impossible to extract finite Casimir energies, so that it has been
suspected that the most divergent term, the so-called bulk energy
contribution, could give rise to the necessary large, virtual
photonic, energy needed.  Dynamically, the difficulty has seemed to
be the extremely short time scales required, which seem to imply
superluminal velocities.

Schwinger \cite{js} and Carlson et al.~\cite{carlson} simply use the
unregulated difference between the zero-point energy of the medium
and that of the `vacuum' within the bubble for their static estimates.
That difference is evidently
\begin{equation}
E_{\rm bulk}={4\pi a^3\over3}\int{(d{\bf k})\over(2\pi)^3}{1\over 2}k\left(
1-{1\over n}\right),
\label{schball}
\end{equation}
where $n$ is the index of refraction of the medium.  This is quartically
divergent, and with a plausible cutoff indeed can yield the required
$10^6$ optical photons per pulse.  However, as we have seen, even in the
simplest geometries, the naive unregulated zero-point energy bears no
relation to the observable, finite, Casimir energy, either in magnitude
or sign.  Although Schwinger had earlier embarked on a calculation of
the Casimir energy for a dielectric bubble, he abandoned the effort before
its completion \cite{jscassph}.
However, I had considered the complementary situation, that of a dielectric
ball, already in 1980 \cite{ball}.  The extension to a dielectic-diamagnetic
ball, with permittivity $\epsilon'$ and permeability $\mu'$, surrounded by a
medium with permittivity $\epsilon$ and permeability $\mu$, was given in
1995 \cite{generalball}.  Using the methods given here, the following
formula was derived for the Casimir energy:
\begin{equation}
E=-{1\over4\pi a}\int_{-\infty}^\infty dy\,e^{iy\delta}\sum_{l=1}^\infty
(2l+1)x{d\over dx}\ln S_l,
\end{equation}
where
\begin{equation}
S_l=[s_l(x')e'_l(x)-s_l'(x')e_l(x)]^2
-\xi^2[s_l(x')e'_l(x)+s_l'(x')e_l(x)]^2,
\end{equation}
where again the $s_l$ , $e_l$ are the spherical Bessel functions of
imaginary argument, the quantity $\xi$ is
\begin{equation}
\xi={\sqrt{\epsilon'\mu\over\epsilon\mu'}-1\over\sqrt{\epsilon'\mu\over
\epsilon\mu'}+1},
\end{equation}
the time-splitting parameter is now denoted by $\delta$,
and \begin{equation}
x=|y|\sqrt{\epsilon\mu},\quad x'=|y|\sqrt{\epsilon'\mu'}.
\end{equation}
It is easy to check that this result reduces to that for a perfectly
conducting spherical shell if we set the speed of light inside and out
the same, $\sqrt{\epsilon\mu}=\sqrt{\epsilon'\mu'}$, as well as set
$\xi=1$.  However, if the speed of light is different in the two
regions, the result is no longer finite, but quartically divergent,
and indeed the Schwinger result (\ref{schball}) follows for that
leading divergent term.

But a divergence is not an answer.  This bulk term in fact was present
in the simpler geometry of the parallel dielectric slabs, treated
in Sec.~3,  where it
was removed as a change in the volume energy of the system.  In other
words, the term to be removed is a contribution to the self energy of
the material, since it renormalizes the phenomenologically described
bulk energy of the substance.  This issue is discussed in detail in
Ref.~\cite{nobulk}.  Mathematically, this term is removed by subtracting
off the uniform medium part of the Green's function in each region, just
as we did in Sec.~4.1.  We are then left with a cubically divergent
energy.

To simplify the subsequent mathematics, let us henceforward assume that
the media are dilute and nonmagnetic, $\epsilon-1\ll1$, $\epsilon'-1\ll1$,
and $\mu=\mu'=1$.  This should be sufficiently good for the qualitative
analysis of the situation.  Then the subtracted energy has the
divergence structure
\begin{equation}
E\sim-{(\epsilon'-\epsilon)^2\over4a}{1\over\delta^3}.
\end{equation}
Because the dimensionless time-splitting cutoff here has the structure
$\delta=\tau/a$, this divergence has the form of a contribution to the
surface tension, which therefore should also be removed by renormalization
of a phenomenological parameter.  Then, we are left with a finite, observable
remainder.

Previously \cite{ball,generalball,nobulk} I had kept merely the leading
term in the uniform asymptotic expansion in estimating this finite result,
believing, as with the perfectly conducting sphere, that that approximation
should be quite accurate.  But Brevik and Marachevsky calculated the next
terms in that expansion and proved me wrong \cite{brevik}.
So jointly we then computed the Casimir energy exactly for a dilute
dielectric sphere \cite{vdw}.  That calculation can be done either by the
point-split cutoff method sketched here, or directly by the zeta function
trick, which swallows all the divergences.  In the latter approach, we
add and subtract the first two leading uniform asymptotic approximants
\begin{equation}
E={(\epsilon-1)^2\over64a}\sum_{l=1}^\infty\left\{\nu^2-{65\over128}\right\}
+E_R\quad (\nu=l+1/2\to\infty),
\end{equation}
where the remainder is the difference between the exact expression and
the asymptotic one.  The $l$ sums here are evaluated according to the
zeta function recipe,
\begin{equation}
\sum_{l=0}^\infty\nu^{-s}=(2^s-1)\zeta(s).
\end{equation}
Keeping only the first term in the asymptotic series gives the result
that I previously quoted, $E_1\sim-(\epsilon-1)^2/(256a)$, while including
the first two terms reverses the sign but hardly changes the magnitude
\cite{brevik}: $E_2\sim(\epsilon-1)^2 33/(8192a)$.  Stopping here would
give a 15\% overestimate; including the remainder gives the 
result\footnote{Immediately after I had completed this calculation,
I received a manuscript from Barton with the same result \cite{barton}
calculated by a more elementary method using ordinary perturbation theory.}
\cite{vdw}
\begin{equation}
E=(\epsilon-1)^2{0.004767\over a},
\label{dilutesphcas}
\end{equation}
which is approximated to better than 2\% by keeping the third term
in the asymptotic expansion \cite{brevik}.

What is most remarkable about this result is that it coincides with
the van der Waals energy calculated two years earlier for this nontrivial
geometry \cite{nobulk}.  That is, starting from the Casimir-Polder
potential \cite{CasimirandPolder}  (\ref{caspol}) we summed the
pairwise potentials between molecules making up the media.  A sensible
way to regulate this calculation is dimensional continuation, similar to
that described in Sec.~6.  That  is, we evaluate the
integral
\begin{equation}
E_{\rm vdW}=-{23\over8\pi}\alpha^2N^2\int d^Dr\,d^Dr'(r^2+r^{\prime2}
-2rr'\cos\theta)^{-\gamma/2},
\end{equation}
where $\theta$ is the angle between $\bf r$ and $\bf r'$, by first
regarding $D>\gamma$ so the integral exists.  The integral may be
done exactly in terms of gamma functions, which when continued to
$D=3$, $\gamma=7$ yields \cite{nobulk}
\begin{equation}
E_{\rm vdW}={23\over1536\pi a}(\epsilon-1)^2.
\end{equation}
This attractive result is numerically identical with the Casimir result
(\ref{dilutesphcas}).
(Incidentally, the dilute Casimir effect for a cylinder, or equivalently,
the van der Waals energy, vanishes \cite{nest,romeo2}.)

Thus, there is no longer any doubt that the Casimir effect is coincident
with van der Waals forces.  The fact that the latter is physically
obviously proportional to $(\epsilon-1)^2$ (which had been already
emphasized in Ref.~\cite{ball}) seems convincing proof that the expression
(\ref{schball}) is incorrect.  In turn, this seems to demonstrate
the irrelevance of the Casimir effect to the light production mechanism
in sonoluminescence.  This is because nearly certainly the adiabatic
approximation is valid (because the flash time is about $10^{-11}$ s,
far longer that the characteristic optical time scale of $10^{-15}$ s.
(Thus the instantaneous approximation of Ref.~\cite{visser}, which on
the face seems to require superluminal velocities, is not credible.)
For details, see Ref.~\cite{sono3}.

\section{Conclusions}
\setcounter{equation}{0}%

In this survey of the Casimir effect, I have been very selective.  Emphasis
has been upon my own work, in part 
because of its greater familiarity.  However,
that emphasis also reflects my bias toward the superiority of Green's
function techniques over, say, zeta function methods, which may reach
the correct answer by defining away physically interesting divergent
contributions.  The first six sections of this report treat unambiguously
finite Casimir forces, about which there can be no argument.  That is not
to say there is not yet work to be done there, for it remains true
that even the sign of
the Casimir force is unpredictable without detailed mathematical
calculation.

Section 7 is far more speculative, in that when interior and exterior
modes are divorced, even so mildly as through a different speed of light,
the Casimir effect is no longer finite.  Removal of that divergence
requires some sort of renormalization, which remains controversial.
Thus, it is still possible to disagree about the possible relevance of
the Casimir effect to sonoluminescence, although I, of course, have
a definite (negative) opinion on the matter.  Recent developments in
this domain of the Casimir effect suggest that earlier calculations
of the Casimir effect in the bag model of hadrons be re-examined
\cite{milton80a,milton80,towardfinite}.

Due to lack of time and space, many other interesting topics had to
be omitted.  For example, the Casimir effect likely plays a role
in the compactification of higher dimensional unified models \cite{kk}.
It may also be relevant in 2-dimensional condensed matter systems \cite{ng}.
Most important are the small, but conceptually vital, inclusion of
radiative corrections, for everything discussed here has been at the
one-loop level \cite{radcor}.  For these, and many other applications
and technical developments, the reader is referred to the original
literature, or to my forthcoming book on the subject \cite{casbook}.

\section*{Acknowledgements}
I am very grateful to Choonkyu Lee for inviting me to the 17th 
Symposium on Theoretical Physics at Seoul National University,
and for his gracious hospitality, and to all the participants in the
Symposium for their interest and encouragement.  This work was supported
in part by the U.S. Department of Energy.

\end{document}